\documentclass[aps,prd,notitlepage,nofootinbib,preprintnumbers,superscriptaddress,10pt]{revtex4-2}
\usepackage{here}
\usepackage{graphicx}
\usepackage{amsmath,amsthm,amssymb}
\usepackage{bm}
\usepackage{color}

\usepackage{amsfonts}
\usepackage{dcolumn}
\usepackage{hyperref}
\allowdisplaybreaks[1]
\usepackage{stackengine}

\newcommand{\be}{\begin{eqnarray}}
\newcommand{\ee}{\end{eqnarray}}

\newcommand{\bem}{\begin{bmatrix}}
\newcommand{\eem}{\end{bmatrix}}

\newcommand{\mF}{\mathcal{F}}

\allowdisplaybreaks
\begin{document}
\title{Black hole thermodynamics in generalized Proca theories}

\author{Masato Minamitsuji}
\affiliation{Faculty of Health Sciences, Butsuryo College of Osaka, Sakai, 593-8328, Osaka, Japan}
\affiliation{
Centro de Astrof\'{\i}sica e Gravita\c c\~ao - CENTRA, Departamento de F\'{\i}sica, Instituto Superior T\'ecnico - IST, Universidade de Lisboa - UL, Av.~Rovisco Pais 1, 1049-001 Lisboa, Portugal}

\author{Kei-ichi Maeda}
\affiliation{
Department of Physics, Waseda University, 3-4-1 Okubo, Shinjuku, Tokyo 169-8555, Japan}
\affiliation{Center for Gravitational Physics and Quantum Information, Yukawa Institute for Theoretical Physics, Kyoto University, 606-8502, Kyoto, Japan}

\begin{abstract}
We investigate thermodynamics of static and spherically symmetric black holes (BHs) in generalized Proca (GP) theories by applying the Iyer-Wald prescription. BH solutions in GP theories are divided into the two classes. The first class corresponds to the solutions obtained by the direct promotion of the BH solutions in shift-symmetric Horndeski theories, while the second class consists of the solutions which are obtained only in GP theories and contain a nonzero electromagnetic field. For BH solutions in the first class, we confirm that the BH entropy and its thermodynamic mass remain the same as those in the counterpart solution in shift-symmetric Horndeski theories. We also calculate the thermodynamical variables of the several static and spherically-symmetric BH solutions in the second class, and 
investigate the thermodynamical stability when there exist two BH solutions. 
\end{abstract}

\date{\today}


\maketitle

\section{Introduction}
\label{introsec}

General Relativity (GR) is the unique gravitational theory in four dimensions which only contains two degrees of freedom (DOFs) of metric and preserves the Lorentz symmetry~\cite{Lovelock:1972vz}.
While GR has been tested by the local experiments as well as the astrophysical probes \cite{Will:2014kxa}, the standard cosmological model based on GR has been plagued by tensions of today's measurements~\cite{Riess:2019cxk,DiValentino:2021izs}, which has motivated us to study gravitational theories other than GR~\cite{Sotiriou:2008rp,Clifton:2011jh,Will:2014kxa,Berti:2015itd}.

Robust modifications to GR are described by scalar-tensor (ST) theories that possess a scalar field ($\phi$) DOF as well as the metric tensor ($g_{\mu\nu}$) DOFs~\cite{Fujii:2003pa}.
The framework of ST theories have been extensively generalized since the (re)discovery of Horndeski theories \cite{Horndeski:1974wa,Deffayet:2009wt,Kobayashi:2011nu}, which are known as the most general ST theories with second-order equations of motion, despite the existence of higher-derivative interactions of the metric and scalar field. 
Horndeski theories are characterized by the four independent coupling functions $G_{2,3,4,5} (\phi,X)$, where $X:=-(1/2)g^{\mu\nu}\nabla_\mu\phi\nabla_\nu\phi$ represents the canonical kinetic term of the scalar field with $\nabla_\mu$ being the covariant derivative associated with the metric $g_{\mu\nu}$.

Within the framework of shift-symmetric Horndeski theories that are invariant under the constant shift transformation~$\phi \to \phi+c$, where the functions~$G_{2,3,4,5}$ depend only on $X$, Ref.~\cite{Hui:2012qt} showed that a no-hair result of static and spherically symmetric BH solutions holds under several hypotheses. If one violates at least one of them, it is possible to realize BH solutions endowed with nontrivial scalar field profile.
For a scalar field with the linear dependence on time~$t$ of the form $\phi=qt+\psi(r)$ with $q$ being constant, there exist the stealth Schwarzschild solution~\cite{Babichev:2013cya,Kobayashi:2014eva,Babichev:2015rva} and the BH solution with asymptotically (anti-) de Sitter~[(A)dS] spacetimes \cite{Babichev:2013cya,Babichev:2016fbg}
\footnote{By `a stealth solution', we mean a solution where the spacetime geometry remains that of a GR BH solution, i.e., the Schwarzschild or Kerr solution, whie the profile of the scalar field becomes nontrivial.}
If the asymptotic flatness is abandoned, the linear term of $X$ in $G_4$ gives rise to the exact hairy BH solution with an asymptotic geometry mimicking the Schwarzschild--AdS spacetime~\cite{Rinaldi:2012vy,Anabalon:2013oea,Minamitsuji:2013ura,Cisterna:2014nua}.
For the coupling $G_5\propto \ln |X|$, which is equivalent to the linear coupling to the Gauss-Bonnet (GB) term, there exists the hairy asymptotically-flat hairy BH solution~\cite{Sotiriou:2013qea,Sotiriou:2014pfa}. 
There also exists the asymptotically-flat BH solution in the model where $G_4(X)\supset (-X)^{1/2}$~\cite{Babichev:2017guv}.
Hairy BH solutions also exist for non-shift-symmetric couplings to the GB term~\cite{Kanti:1995vq,Kanti:1997br,Torii:1996yi,Silva:2017uqg,Doneva:2017bvd,Antoniou:2017acq,Blazquez-Salcedo:2018jnn,Minamitsuji:2018xde,Silva:2018qhn,Cunha:2019dwb,Konoplya:2019fpy,Doneva:2021dqn,East:2021bqk,Julie:2022huo,Doneva:2020nbb,Doneva:2022yqu,Dima:2020yac,Herdeiro:2020wei,Lai:2023gwe}.

The linear stability of static and spherically symmetric BH solutions in Horndeski theories have been studied in the literature, e.g., \cite{Kobayashi:2012kh,Kobayashi:2014wsa,Kase:2021mix,Minamitsuji:2022mlv,Minamitsuji:2022vbi}. 
However, in the models where the scalar field linearly depends on time, e.g., the stealth Schwarzschild solution \cite{Babichev:2013cya,Babichev:2016fbg}, the standard linear perturbation analysis cannot be applied because the perturbations become infinitely strongly coupled \cite{deRham:2019gha,Motohashi:2019ymr}.
For such solutions, there could be several ways to clarify their stability.
One of them is to investigate their thermodynamical stability.
When one has two BH solutions in a given gravitational theory, one can argue that the BH solution with smaller entropy is more unstable than one with larger entropy. 
Wald's entropy formula \cite{Wald:1993nt} may not be directly applicable to Horndeski theories in the presence of derivative interactions between the scalar field and the spacetime curvature \cite{Feng:2015oea,Hajian:2020dcq}.
This is because the apparent dependence of the action on the Riemann curvature of the spacetime may be modified before and after a partial integration and there may exist an additional contribution from higher-derivative coupling of a scalar field.

In our former work \cite{Minamitsuji:2023nvh}, following the original 
prescription by Iyer and Wald~\cite{Iyer:1994ys}, we have constructed  thermodynamical variable for static and spherically symmetric BH solutions in Horndeski theories.
Because of the four-dimensional diffeomorphism invariance, there exists the associated Noether charge.
As shown by Iyer and Wald, the variation of the Hamiltonian is given by that of the Noether charge evaluated on the boundaries of the Cauchy surface, i.e., in our case, the BH event horizon and spatial infinity \cite{Iyer:1994ys}.
The conservation of the total Hamiltonian of the BH system reproduces the first law of BH thermodynamics.
The Iyer-Wald prescription had been applied for BH solutions in a particular class of Horndeski theories in Refs.~\cite{Feng:2015oea,Hajian:2020dcq}.
It had also been applied to the planar BH solutions in some classes of Horndeski theories~\cite{Bravo-Gaete:2021hlc}.
Ref. \cite{Minamitsuji:2023nvh} analyzed the whole Horndeski theories and was able to apply all classes of the static and spherically symmetric BH solutions including those with the linearly time-dependent scalar field in shift-symmetric theories \cite{Babichev:2013cya,Babichev:2016fbg}.

Horndeski theories have been straightforwardly extended to the vector-tensor (VT) theories which admit the second-order equations of motion despite the presence of derivative interactions. 
These theories are called generalized Proca (GP) theories \cite{DeRham:2014wnv,DeFelice:2016yws,DeFelice:2016cri}.
The interactions in the purely longitudinal sector in GP theories were obtained by promoting the derivative of the scalar field $\nabla_\mu\phi$ in shift symmetric Horndeski theories to the vector field $A_\mu$.
Along such an extension, $X$ and $\Box\phi$ in ST theories can be generalized to $Y=-(1/2)g^{\mu\nu}A_\mu A_\nu$ and $\nabla^\mu A_\mu$ in VT theories.
Beside them, there are also derivative interactions containing the electromagnetic polarizations which are given by the contractions including the field strength of the vector field $F_{\mu\nu}=\partial_\mu A_\nu-\partial_\nu A_\mu$.
Up to the generalized quartic-order interactions, we find the explicit form of the action of GP theories as shown later.

BH solutions with nontrivial profiles of the vector field in GP theories have been obtained in Refs.~\cite{Geng:2015kvs,Chagoya:2016aar,Minamitsuji:2016ydr,Cisterna:2016nwq,Fan:2016jnz,Heisenberg:2017xda,Heisenberg:2017hwb,Minamitsuji:2017aan,Babichev:2017rti,Chagoya:2017ojn,Fan:2017bka,Chagoya:2023ddb}.
The stealth Schwarzschild solutions and the Schwarzschild-(A)dS solutions with the scalar field $\phi=qt+\psi(r)$ in shift symmetric Horndeski theories \cite{Babichev:2013cya,Kobayashi:2014eva,Babichev:2015rva,Babichev:2016fbg} have been straightforwardly promoted to those with $A_t=q$ and $A_r=\psi'(r)$ in GP theories~\cite{Chagoya:2016aar,Minamitsuji:2016ydr,Fan:2016jnz,Heisenberg:2017xda,Heisenberg:2017hwb,Minamitsuji:2017aan,Babichev:2017rti}.
Besides the straightforward promotion of the stealth Schwarzschild solutions and the Schwarzschild-(A)dS BH solutions to GP theories, there also exist the stealth Schwarzschild solutions and the Schwarzschild-(A)dS solutions with the nonzero electric charge, if the coupling parameter is chosen to the certain value~\cite{Chagoya:2016aar,Minamitsuji:2016ydr,Heisenberg:2017xda,Heisenberg:2017hwb,Minamitsuji:2017aan}.
There are also the other BH solutions with nontrivial vector hair in various classes of GP theories~\cite{Minamitsuji:2016ydr,Minamitsuji:2017aan,Chagoya:2017ojn,Fan:2017bka}.
The linear stability of the hairy BH solutions in GP theories has been investigated in Refs.~\cite{Kase:2018voo,Garcia-Saenz:2021uyv}.
In the present work, by applying the 
prescription by Iyer and Wald and extending the analysis in Ref.~\cite{Minamitsuji:2023nvh}, we will investigate thermodynamical properties of the static and spherically symmetric BH solutions in GP theories.
In the case that Horndeski and GP theories share the same BH solutions, we will carefully check whether the entropy and the thermodynamic mass of BHs remain unchanged between the two descriptions in ST and VT theories.
We will also discuss the thermodynamic properties for several exact BH solutions which are generic to GP theories.
While Horndeski and GP theories do not satisfy the constraint on the speed of gravitational waves \cite{TheLIGOScientific:2017qsa,Baker:2017hug,Ezquiaga:2017ekz}, $c_{\rm GW}=c$, ST and VT theories beyond them \cite{Langlois:2015cwa,BenAchour:2016fzp,Kimura:2016rzw}, contain the BH solutions with very similar properties \cite{Takahashi:2019oxz,deRham:2019gha,Takahashi:2021bml,Minamitsuji:2021gcq}.
Thus, in order to understand thermodynamic stability of BHs in Horndeski and GP  theories will give a direct hint for BH thermodynamics of these extended theories.

The paper is constructed as follows:
In Sec. \ref{sec2}, we review GP theories, and derive the boundary current and the Noether charge associated with the diffeomorphism invariance.
In Sec. \ref{sec3}, we present thermodynamic quantities of static and spherically symmetric BHs in GP theories in terms of the variation of the Noether charge associated with the diffeomorphism invariance, which are evaluated at the boundaries of the Cauchy surface.
In Sec. \ref{sec4}, we investigate BH thermodynamics of the stealth Schwarzschild solutions and the Schwarzschild-(A)dS solutions without and with the electric charge.
We also compare their entropies with the peculiar solution in the same theory.
In Sec. \ref{sec5}, we study the thermodynamical stability of the other BH solutions in GP theories including the generalized quadratic- and cubic-order interactions.
In Sec. \ref{sec6}, we close the paper after giving the brief summary and conclusion.

\section{Generalized Proca theories and Noether charge for diffeomorphism}
\label{sec2}

\subsection{Generalized Proca theories}

We consider the action for GP theories which is given by 
\be
\label{action}
S_{\rm GP}:=\sum_{i=2}^4S_{(i)},
\ee
where the generalized quadratic-, cubic-, and quartic-order interactions are, respectively, defined by 
\be
S_{(2)}
&=&
\int d^4x
\sqrt{-g}
G_2
\left(
{\cal F},\tilde{\cal F},Y
\right),
\label{l2}
\\
S_{(3)}
&=&
\int d^4x
\sqrt{-g}
\left(
-G_3(Y) \nabla^\mu A_\mu
\right),
\label{l3}
\\
\label{quartic}
S_{(4)}
&=&
\int 
d^4x
\sqrt{-g} 
\left\{
G_4 (Y)R
+
G_{4Y} (Y)
\left[
\left(\nabla^\mu A_\mu\right)^2
+
c_2(Y) \nabla^\rho A^\sigma \nabla_\rho A_\sigma
-
\left(
1+c_2(Y)
\right) 
\nabla^\rho A^\sigma \nabla_\sigma A_\rho
\right]
\right\}.
\label{l4}
\ee
Here, $g_{\mu\nu}$ is the spacetime metric tensor, $\nabla_\mu$ represents the covariant derivative associated with $g_{\mu\nu}$, $R$ is the Ricci scalar associated with $g_{\mu\nu}$, $A_\mu$ is the vector field, $F_{\mu\nu}:=\nabla_\mu A_\nu-\nabla_\nu A_\mu$ represents the electromagnetic field strength of the vector field $A_\mu$, and 
\be
\label{def_fx}
{\cal F}:=-\frac{1}{4}g^{\alpha\beta} g^{\rho\sigma} F_{\alpha\rho}F_{\beta\sigma},
\qquad 
Y:=-\frac{1}{2}g^{\mu\nu}A_\mu A_\nu,
\qquad
\tilde{\cal F}:=-\frac{1}{4}{\tilde F}^{\mu\nu} F_{\mu\nu},
\ee
with ${\tilde F}^{\mu\nu}:=\frac{1}{2} \epsilon^{\mu\nu\alpha\beta}F_{\alpha\beta}$ being the 2-form field strength dual to $F_{\mu\nu}$. $G_2\left({\cal F},\tilde{\cal F},Y \right)$ is the free function of the given arguments. Similarly, $G_{3}(Y)$, $G_4(Y)$, and $c_2(Y)$ are the free functions of $Y$, respectively. 
The equations of motion of GP theories are given by the second-order differential equations \cite{DeRham:2014wnv,DeFelice:2016yws,DeFelice:2016cri}.

Using the St\"{u}ckelberg trick $A_\mu\to A_{\mu}+\nabla_\mu \varphi$, the $U(1)$ symmetry under the transformation $\varphi\to \varphi+\chi$ and $A_\mu\to A_\mu-\nabla_\mu \chi$ is restored. In the absence of the electromagnetic polarizations, $Y\to X=-\frac{1}{2}g^{\mu\nu}\nabla_\mu \varphi \nabla_\nu\varphi$, $\nabla^\mu A_\mu \to \Box\varphi= g^{\mu\nu}\nabla_\mu \nabla_\nu \varphi$, and then GP interactions \eqref{l2}-\eqref{l4} reduce to
\be
\label{quadratic2}
S_{(2)} &=& \int d^4x \sqrt{-g} {\tilde G}_2 \left(X\right),
\qquad
{\tilde G}_{2}(X):= G_2\left(
0,0,X\right),
\\
\label{cubic2}
S_{(3)}
&=&
\int d^4x \sqrt{-g} \left(-G_3(X) \Box\varphi \right),
\\
\label{quartic2}
S_{(4)}
&=&
\int d^4x \sqrt{-g} \left\{G_4 (X)R
+G_{4X} (X) \left[
\left(\Box\varphi\right)^2
-
\nabla^\mu \nabla^\nu \varphi
\nabla_\mu \nabla_\nu \varphi
\right]
\right\},
\ee
which correspond to shift-symmetric Horndeski theories with the metric $g_{\mu\nu}$ and the `scalar field' $\varphi$, up to the generalized quartic-order interaction. Thus, in the absence of the electromagnetic modes, any solution in GP theories with $A_\mu$ reduces to that in shift-symmetric Horndeski theories with the scalar field $\varphi$.

\subsection{Equations of motion and Boundary current}

The variation of the action \eqref{action} is given by 
\be
\delta S
=
\int d^4x
\sqrt{-g}
\left(
E_{\mu\nu} 
\delta g^{\mu\nu}
+
E_{A_\mu}
\delta A^\mu
+
\nabla_\mu J^\mu
\right),
\ee
where $E_{\mu\nu}=0$ and $E_{A_\mu}=0$ are the Euler-Lagrange (EL) equations for the metric and vector field,  respectively. The boundary term is given by the total derivative of the covariant current
\be
\label{boundary_current}
J^\mu=\sum_{i=2}^4J_{(i)}^\mu,
\ee
with the contribution from each order of GP interactions
\be
\label{j2}
J_{(2)}^\rho&:=&
-G_{2{\cal F}}F^{\rho\sigma}\delta A_\sigma-G_{2\tilde{\cal F}} {\tilde F}^{\rho\sigma}\delta A_\sigma,
\\
\label{j3}
J_{(3)}^\rho
&=&
G_3(Y)
\left(
g^{\rho \mu} A^\nu
-
\frac{1}{2}
A^\rho
g^{\mu\nu}
\right)
\delta
g_{\mu\nu}
-
G_3(Y)
g^{\rho\nu}
\delta A_\nu,
\\
\label{j4}
J^\rho_{(4)}
&=&
G_4(Y)
\left[
\nabla^\nu 
\left(g^{\rho\sigma}\delta g_{\sigma\nu}\right)
-
\nabla^\rho
\left(g^{\nu\sigma}\delta g_{\sigma\nu}\right)
\right]
+
G_{4Y}(Y)
A^\alpha
\left[
g^{\nu\rho} 
\nabla^\sigma 
A_\alpha
-
g^{\sigma\nu}
\nabla^\rho
A_\alpha
\right]
\delta g_{\sigma\nu}
\nonumber
\\
&+&
G_{4Y} (Y) \nabla^\nu A_\nu
\left[
2g^{\rho\sigma} \delta A_\sigma
-
2g^{\rho\sigma} A^\nu \delta g_{\nu\sigma}
+
A^\rho g^{\nu\sigma} \delta g_{\nu\sigma}
\right]
\nonumber\\
&-&
2
G_{4Y} (Y)
\left[
\left(\nabla^\sigma A^\rho\right) \delta A_\sigma
-
A^\nu \nabla^{(\sigma} A^{\rho)}
\delta g_{\nu\sigma}
+
\frac{1}{2}
A^\rho
\nabla^\nu A^\sigma
\delta g_{\sigma\nu}
\right]
+
2
c_2(Y)
G_{4Y}(Y)
F^{\rho\sigma}
\delta A_\sigma,
\ee
where we have defined $G_{2\mF}:=\frac{\partial G_2}{\partial \mF}$, $G_{2\tilde{\cal F}}:=\frac{\partial G_2}{\partial \tilde{\cal F}}$ and $G_{iY}:=\frac{\partial G_i}{\partial Y}$ ($i=2,3,4$).  We also define the 3-form dual to the current $J^\mu$, Eq. \eqref{boundary_current}, by 
\be
\label{dual_3form}
{\Theta}_{\alpha\beta\gamma}
&:=&
J^\mu
\varepsilon_{\mu\alpha\beta\gamma}
=
{\varepsilon}_{\alpha\beta\mu \gamma}
J^\mu
=
\sum_{i=2}^4
{\varepsilon}_{\alpha\beta\mu \gamma}
J_{(i)}^\mu.
\ee

In Sec. \ref{sec4}, we will focus on the class of GP theories with the following choice of the functions
\be
\label{particular_theory}
G_2
&=&
{\cal F}+2m^2 Y-\frac{\Lambda}{8\pi G},
\qquad
G_4(Y)=\frac{1}{16\pi G}+\beta Y,
\qquad 
G_3(Y)
=
c_2(Y)
=0,
\ee
which is equivalent to the VT theory with the nonminimal coupling of the vector field to the Einstein tensor $G^{\mu\nu}$
\be
\label{action_nmc}
S_{\rm NMC}
=
\int d^4 x\sqrt{-g}
\left(
\frac{1}{16\pi G}
\left(
R-2\Lambda
\right)
+{\cal F}
+ 
2m^2Y
+
\beta G^{\mu\nu}A_\mu A_\nu
\right).
\ee
Within the theory \eqref{particular_theory},
the stealth Schwarzschild solutions and the Schwarzschild-(A)dS solutions have been obtained in this GP theory~\cite{Chagoya:2016aar,Minamitsuji:2016ydr,Heisenberg:2017xda,Heisenberg:2017hwb,Minamitsuji:2017aan} for the first time.
We note that the current from the variation of the action \eqref{action_nmc} is given by
\be
\label{current_nmc}
J_{\rm NMC}^\rho
&:=&
-F^{\rho\sigma} \delta A_\sigma
+
\frac{1}{16\pi G}
\left[
\nabla^\nu
\left(
g^{\rho\beta}\delta g_{\beta\nu}
\right)
-
\nabla^\rho
\left(
g^{\beta\nu}\delta g_{\beta\nu}
\right)
\right]
\nonumber\\
&+&
\beta
\left[
A^\alpha A^\beta g^{\rho\sigma} 
\nabla_\beta \delta g_{\sigma\alpha}
-
\frac{1}{2}
A^\sigma A^\alpha
\nabla^\rho \delta g_{\sigma\alpha}
-
\frac{1}{2}
A^\beta A^\rho g^{\alpha\sigma}
\nabla_\beta \delta g_{\sigma\alpha}
-
\frac{1}{2}
A^\mu A_\mu
\left(
\nabla^\alpha
\left(
g^{\rho\sigma}
\delta g_{\sigma\alpha}
\right)
-
\nabla^\rho
\left(
g^{\alpha\sigma}
\delta g_{\sigma\alpha}
\right)
\right)
\right]
\nonumber
\\
&-&
\beta
\Big[
\nabla^\sigma
\left(
A^\rho A^\alpha
\right)
-
\frac{1}{2}
\nabla^\rho
\left(
A^\sigma A^\alpha
\right)
-
\frac{1}{2}
\nabla^\nu
\left(
A^\rho A_\nu
\right)
g^{\sigma\alpha}
-
\frac{1}{2}
\nabla^\sigma
(A^\nu A_\nu)
g^{\rho\alpha}
+
\frac{1}{2}
\nabla^\rho
(A^\nu A_\nu)
g^{\alpha\sigma}
\Big]
\delta 
g_{\sigma\alpha}.
\ee
As we will see later, this is not exactly the same as the current (\ref{boundary_current}) with the condition (\ref{particular_theory}), but the thermodynamical quantities become the same.

\subsection{Noether charge for the differmorphism invariance}

Since GP theories~\eqref{action} with Eqs. \eqref{l2}-\eqref{l4} are invariant under the diffeomorphism transformation, $x^\mu\to x^\mu+\xi^\mu (x^\mu)$, there exists the associated Noether current. 
Under the diffeomorphism transformation, the variations of the metric and the vector field are, respectively, given by 
\be
\label{diffeo}
\delta_\xi g_{\mu\nu}
&=&
\nabla_\mu \xi_\nu
+
\nabla_\nu \xi_\mu,
\qquad 
\delta_\xi A_\mu
=
\xi^\sigma
\nabla_\sigma A_\mu
+
A_\sigma
\nabla_\mu
\xi^\sigma,
\ee
and with use of the EL equations, we obtain the conserved Noether current for the differmorphism invariance
\be
\label{jmu}
J_{(\xi)}^\mu 
-
\xi^{\mu} 
{\cal L}
=
2
\nabla_\nu K^{[\nu\mu]}_{(\xi)}
=
2
\sum_{i=2}^4
\nabla_\nu
 K^{[\nu\mu]}_{(i,\xi)},
\ee
where the contribution from each order of GP interactions is given by 
\be
K_{(2,\xi)}^{\mu\nu}
&:=&
\frac{1}{2}
\left(
G_{2{\cal F}}
F^{\mu\nu}
+
G_{2\tilde{\cal F}}
\tilde{F}^{\mu\nu}
\right),
\\
K_{(3,\xi)}^{\mu\nu}
&:=&
-
G_3(Y)
\xi^{\mu} A^{\nu},
\\
K_{(4,\xi)}^{\mu\nu}
&=&
G_4(Y) \nabla^{\mu}\xi^{\nu}
+
2G_{4Y}(Y)A^\alpha \left(\nabla^{\mu}A_\alpha \right)\xi^{\nu}
-
2G_{4Y}(Y)\left (\nabla^\alpha A_\alpha \right) A^{\mu}\xi^{\nu}
\nonumber
\\
&-&
\frac{1}{2}
\left(
1+
2c_2(Y)\right)
 G_{4Y}(Y) F^{\mu\nu}A^\alpha \xi_\alpha
+
G_{4Y}(Y)
\left(
A^{\mu}\nabla^{\nu} A^\alpha
+
A^{\mu}\nabla^\alpha A^{\nu}
\right)
\xi_\alpha.
\ee
This is because the right-hand of Eq. \eqref{jmu} is given by the 
divergence
of the anti-symmetric rank-2 tensor 
$K^{[\mu\nu]}_{(\xi)}$.
Thus, $K^{[\mu\nu]}_{(\xi)}$ can be interpreted as the Noether charge for the diffeomorphism invariance.

Similarly, for the theory with the nonminimal coupling to the Einstein tensor \eqref{action_nmc} (see Sec. \ref{sec4}), the Noether charge is given by 
\be
K_{{\rm NMC}(\xi)}^{\mu\nu}
&=&
\frac{1}{2} F^{\mu\nu}A_\sigma \xi^\sigma
+
\frac{1}{16\pi G}
\nabla^\mu \xi^\nu
\nonumber\\
&+&
\beta
\Big[
A^\beta A^\mu \nabla_\beta \xi^\nu
-
\frac{1}{2} 
\left(A^\rho A_\rho\right)
\nabla^\mu \xi^\nu
+
2A^\alpha (\nabla^\mu A_\alpha)\xi^\nu
-
\left(\nabla_\alpha A^\alpha\right)
A^\mu \xi^\nu
-
A^\alpha
\left(\nabla_\alpha A^\mu \right) \xi^\nu
\nonumber\\
&&
-
\frac{1}{2}
F^{\mu\nu}
A^\sigma \xi_\sigma
+
A^\mu \left(\nabla^\nu A^\sigma\right)A_\sigma
\Big].
\ee
We then define the 2-form tensor dual to the Noether charge $K_{(\xi)}^{[\mu\nu]}$ \cite{Peng:2015yjx}
\be
\label{def_q}
{Q}_{(\xi)\alpha\beta}
:=
-\epsilon_{\alpha\beta\mu\nu}
\sum_{i=2}^4 K^{\mu\nu}_{(i,\xi)}
=
\sum_{i=2}^4
{Q}^{(i)}_{(\xi)\alpha\beta}.
\ee
We also define the 2-form tensor where the first index of ${\Theta}_{\nu\alpha\beta}$ defined in Eq.~\eqref{dual_3form} is contracted by the infinitesimal differmorphism transformation $\xi^\nu$ by 
\be
\label{itheta}
i_\xi {\Theta}_{\alpha\beta}
&:=&
\xi^\nu
{\Theta}_{\nu\alpha\beta}
=
\varepsilon_{\alpha\beta\mu\nu}
J^\mu \xi^\nu.
\ee
We now consider the variation of the dual Noether charge with respect to the physically independent charges in the solution
subtracted by Eq.~\eqref{itheta}
\be
\label{deltaQ}
\delta{Q}_{(\xi)\alpha\beta}
-
i_\xi {\Theta}_{\alpha\beta}
=
-
\sum_{i=2}^4
\left(
\delta
\left(
\varepsilon_{\alpha\beta\mu\nu}
K^{\mu\nu}_{(i)(\xi)}
\right)
+
\varepsilon_{\alpha\beta\mu\nu}
J_{(i)}^\mu \xi^\nu
\right).
\ee
In Refs.~\cite{Wald:1993nt,Iyer:1994ys}, Iyer and Wald showed that the integration of Eq.~\eqref{deltaQ} on the boundaries of the Cauchy surface gives rise to the variation of the total Hamiltonian of a given BH system.

\section{Thermodynamic variables of static and spherically symmetric Black holes}
\label{sec3}

From now on, we focus on the static and spherically symmetric solutions whose metric is written by 
\be
\label{metric}
ds^2
&=&
-h(r) dt^2
+\frac{dr^2}
          {f(r)}
+r^2\gamma_{ab}d\theta^a d\theta^b,
\ee
where $t$ and $r$ are the temporal and radial coordinates, and $\gamma_{ab}d\theta^a d\theta^b :=d\theta^2+\sin^2\theta d\varphi^2$ represents the metric of the unit two-sphere.
We assume that the spacetime contains the event horizon at $r=r_g$ where $h(r_g)=f (r_g)=0$ and $\lim_{r\to r_g}\frac{f(r)}{h(r)}={\rm const}$.
We also assume the following ansatz for the vector field to be compatible with the staticity and spherical symmetry of the spacetime
\be
\label{vector_field}
A_\mu dx^\mu=A_0(r) dt+A_1(r) dr.
\ee
In the case where $h(r)$ and $f(r)$ have several positive roots, we assume that $r_g$ corresponds to the largest positive root with $h'(r_g)>0$ and $f'(r_g)>0$, which does not include the cosmological horizon by definition.

We consider the case that $\xi^\mu$ corresponds to the timelike Killing vector field $\xi^\mu=(1,0,0,0)$.
The variation of the metric and the vector field of a given solution can be written in terms of those of the independent integration constants
\be
\label{variation}
&&
\mathfrak{h}_{tt}
=
-\delta h
=
-\sum_j \frac{\partial h}{\partial c_j}\delta c_j,
\qquad 
\mathfrak{h}_{rr}
=-\frac{\delta f }{f^2}
=
-\frac{1}{f^2}
\sum_j
\frac{\partial f}{\partial c_j}
\delta c_j,
\qquad 
\mathfrak{h}_{ab}
=0,
\nonumber\\
&&
\delta A_0
=
\sum_j
\frac{\partial A_0}{\partial c_j}
\delta c_j,
\qquad 
\delta A_1
=
\sum_j
\frac{\partial A_1}{\partial c_j}
\delta c_j,
\ee
where $c_j$'s are integration constants of a given BH solution including the position of the event horizon $r_g$.
Following Iyer and Wald~\cite{Wald:1993nt,Iyer:1994ys}, with use of Eq.~\eqref{deltaQ} the variation of the Hamiltonian with respect to the integration constants is given by the contributions from the boundaries, i.e., the horizon $r\to r_g$ and the infinity $r\to \infty$ as
\be
\label{hamiltonian}
\delta {\cal H}
:=
\delta{\cal H}_\infty
-
\delta{\cal H}_{H}
&=&
-
\int
d\Omega
\sum_{i=2}^4
\left(
\delta
\left(
r^2
\sqrt{\frac{h}{f}}
K^{[tr]}_{(i)(\xi)}
\right)
+
r^2
\sqrt{\frac{h}{f}}
J_{(i)}^{[t} \xi^{r]}
\right)
\Big|_{r\to\infty}
\nonumber
\\
&&+
\int
d\Omega
\sum_{i=2}^4
\left(
\delta
\left(
r^2
\sqrt{\frac{h}{f}}
K^{[tr]}_{(i)(\xi)}
\right)
+
r^2
\sqrt{\frac{h}{f}}
J_{(i)}^{[t} \xi^{r]}
\right)
\Big|_{r\to r_g},
\ee
where 
$d\Omega:=\sin\theta d\theta d\varphi$ represents the differential of the solid angle
and the subscript $H$ represents the quantities evaluated at the event horizon.
The variation of the Hamiltonian on the event horizon 
is identified with the variation of the BH entropy $S_{\rm GP}$ 
in GP theories 
\be
\label{var_hamilton}
\delta {\cal H}_{H}=T_{\textsf{H}({\rm GP})} \delta S_{\rm GP},
\ee
where $T_{\textsf{H}({\rm GP})}$ represents the Hawking temperature (denoted by $\textsf{H}$) of the given BH solution
\be
T_{\textsf{H}({\rm GP})}:=\frac{\sqrt{h'(r_g)f'(r_g)}}{4\pi}.
\label{hawking_temperature}
\ee
The conservation of the total Hamiltonian $\delta {\cal H}=0$ leads to the first law of BH thermodynamics in GP theories, 
\be
T_{\textsf{H}({\rm GP})} \delta S_{\rm GP}
=\delta {\cal H}_\infty.
\label{1st_law}
\ee
In order to illustrate that the variation of the Hamiltonian is independent of the description of the theory, we consider the class of GP theories \eqref{particular_theory} which is physically equivalent to the theory~\eqref{action_nmc}, which admits the stealth Schwarzschild and  Schwarzschild-(A)dS solutions~\cite{Chagoya:2016aar,Minamitsuji:2016ydr,Heisenberg:2017xda,Heisenberg:2017hwb,Minamitsuji:2017aan}.
By considering the variation of  Eqs. \eqref{metric} and \eqref{vector_field}, we obtain
\be
\label{cq3_f}
&&
-
\delta
\left(
r^2
\sqrt{\frac{h}{f}}
K^{[tr]}_{{\rm GP}(\xi)}
\right)
-
r^2
\sqrt{\frac{h}{f}}
J_{\rm GP}^{[t}\xi^{r]}
\nonumber
\\
&=&
r^2
\sqrt{\frac{h}{f}}
\Big\{
-\frac{\delta f (r)}{8\pi G r}
+
\frac{A_0(r) A_0'(r) f(r)}{2h(r)^2}
\delta h(r)
-
\frac{A_0(r) A_0'(r)}{2h(r)}
\delta f(r)
-
\frac{A_0(r) f(r)}{h(r)}
\delta A_0'(r)
\nonumber\\
&+&
\beta
\left(
-\frac{3A_1(r)^2f(r)}{r}\delta f(r)
-
\frac{4f(r)^2A_1(r)}{r}
\delta A_1(r)
-
\frac{A_0(r)^2}{rh(r)}
\delta f(r)
\right)
\Big\}.
\ee
Although the Noether charges for the two equivalent theories \eqref{particular_theory} and \eqref{action_nmc} do not coincide in general as
\be
\label{diff_Noether}
K^{[tr]}_{{\rm GP}(\xi)}
-K^{[tr]}_{{\rm NMC}(\xi)}
=\frac{2\beta A_1(r)^2 f(r)^2}{r}
-
\frac{\beta A_0(r)^2 f(r) h'(r)}{4h(r)^2}
+
\frac{\beta A_1(r)^2 f(r)^2 h'(r)}{4h(r)}
\,,
\ee
we find that the integrands of the variation of the Hamiltonian \eqref{hamiltonian} for the two equivalent theories \eqref{particular_theory} and \eqref{action_nmc}
coincide as
\be
-
\delta
\left(
r^2
\sqrt{\frac{h}{f}}
K^{[tr]}_{{\rm GP}(\xi)}
\right)
-
r^2
\sqrt{\frac{h}{f}}
J_{\rm GP}^{[t}\xi^{r]}
=
-
\delta
\left(
r^2
\sqrt{\frac{h}{f}}
K^{[tr]}_{{\rm NMC}(\xi)}
\right)
-
r^2
\sqrt{\frac{h}{f}}
J_{\rm NMC}^{[t}\xi^{r]}.
\ee
Thus, physically relevant quantities for BH thermodynamics do not depend on the description of the theory.
The difference between the Noether charges \eqref{diff_Noether} is exactly due to a $\mu$-ambiguity because the difference between the two Noether charges is originated from the difference of the total derivative terms of the Lagrangians  \eqref{particular_theory} and \eqref{action_nmc} as Eq. (41) of Ref.~\cite{Iyer:1994ys}. We note that the $\mu$-ambiguity does not modify the total Hamiltonian~\cite{Iyer:1994ys}.

Before going to explore BH thermodynamics in GP theories, we shall review BH thermodynamics in the $U(1)$-invariant VT theories, i.e., Einstein-Maxwell (EM) and Einstein-Born-Indeld (EBI) theories.

\subsection{Einstein-Maxwell theory}

In this subsection, we consider EM theory with the cosmological constant $\Lambda$,
\be
G_2
=
{\cal F}
-
\frac{\Lambda}{8\pi G}
\qquad
G_4
=
\frac{1}{16\pi G},
\qquad 
G_3
=
c_2
=
0.
\label{em_f}
\ee
As the unique exact static and spherically symmetric BH solution in EM theory \eqref{em_f}, there exists the Reissner-Nordstr\"{o}m (RN)-dS solution,
\be
\label{bcl_solution_f}
f(r)
=
h(r)
&=&
1
-\frac{\Lambda}{3}r^2
+
\frac{1}{r}
\left(
\frac{r_g}{3}
\left(
-3+r_g^2\Lambda
\right)
-
\frac{4\pi G Q^2}
        {r_g}
\right)
+
\frac{4\pi G Q^2}
        {r^2},
\quad
\nonumber\\
A_0(r)
&=&
\frac{Q}{r}
+
q_{\rm EM},
\quad 
A_1(r)
=0,
\ee
which the constant $q_{\rm EM}$ corresponds to the gauge DOF.

The variation of the Hamiltonian on the horizon $r=r_g$ yields
\be
\label{var_horizon_em}
\delta{\cal H}_H
=
\frac{\delta r_g}{2G}
\left[
1
-r_g^2\Lambda
-
\frac{4\pi G Q^2}
       {r_g^2}
\right]
+
4\pi A_0(r_g)\delta Q,
\ee
However, the same gauge-dependent term proportional to $\delta Q$ also appears in the variation of the Hamiltonian at the infinity $r\to\infty$, $\delta {\cal H}_\infty$. 
Thus, in the conservation of the Hamiltonian, this term proportional to $\delta Q$ cancels.
Hence, the differential of the BH entropy is given by 
\be
T_{\textsf{H}({\rm EM})} \delta S_{\rm EM}
=
\frac{\delta r_g}{2G}
\left[
1
-r_g^2\Lambda
-
\frac{4\pi G Q^2}
       {r_g^2}
\right],
\ee
where the Hawking temperature \eqref{hawking_temperature} is given by $T_{\textsf{H}({\rm EM})}=T_0\left(1-r_g^2\Lambda-\frac{4\pi GQ^2} {r_g^2}\right)$ with
\be
T_0:=\frac{1}{4\pi r_g},
\label{T0}
\ee
being the Hawking temperature of the Schwarzschild BH.
Thus, we obtain the BH entropy
\be
S_{\rm EM}=S_0,
\ee
where 
\be
S_0:= \frac{\pi r_g^2}{G},
\label{S0}
\ee 
represents the Bekenstein-Hawking entropy.

The variation of the Hamiltonian at the infinity $r\to \infty$ yields
\be
\label{var_infinity_em}
\delta {\cal H}_\infty
=
\frac{\partial M_{\rm EM}}
        {\partial r_g}
\delta r_g
=
\delta M_{\rm EM}
-
\frac{\partial M_{\rm EM}}
        {\partial Q}
\delta Q
=
\delta M_{\rm EM}
-
\Phi_H
\delta Q,
\ee
where $\Phi_H:=- 4\pi \left(A_0(r\to \infty)-A_0(r_g)\right)=\frac{4\pi Q}{r_g}$ describes the difference in the electric potential between at the infinity $r\to \infty$ and at the horizon $r=r_g$.
The conservation of the Hamiltonian ${\cal H}=0$ yields the first law of thermodynamics for the electrically charged BH,
\be
\label{charged_first_law_em}
T_{\textsf{H}({\rm EM})}
\delta S_{\rm EM}
=
\delta M_{\rm EM}
-
\Phi_{H}
\delta Q.
\ee
Thus, by integrating this, the thermodynamic mass of BHs is given by 
\be
M_{\rm EM}
=
M_0
\left(
1
+
\frac{4\pi G Q^2}
       {r_g^2}
-\frac{\Lambda r_g^2}{3}
\right),
\ee
where
\be
M_0:=\frac{r_g}{2G},
\label{M0}
\ee
represents the thermodynamic mass in the Schwarzschild background, which coincides with the Arnowitt-Deser-Misner (ADM) mass of the BH spacetime.

EM theory \eqref{em_f} also admits the Schwarzschild-(A)dS solution with the trivial vector field
\be
\label{em_gr}
h(r)
=
f(r)
=
1
-
\frac{\Lambda r^2}{3}
-
\frac{r_g}{r}
\left(
1
-
\frac{\Lambda r_g^2}{3}
\right),
\qquad 
A_0(r)
=
A_1(r)
=
0.
\ee
The variation of the Hamiltonian yields $\delta{\cal H}_H=T_{\textsf{H}({\rm GR})} \delta S_{\rm GR}=\frac{\delta r_g}{2G}$,
where the Hawking temperature \eqref{hawking_temperature} is given by $T_{\textsf{H}({\rm GR})}=T_0$.
Thus, we obtain the Bekenstein-Hawking entropy as the BH entropy $S_{\rm GR}=S_0$.
The variation of the Hamiltonian at the infinity $r\to \infty$ yields $\delta  M_{\rm GR}=\frac{\delta r_g}{2G}$.
The thermodynamic mass of the BH is given by $M_{\rm GR}=M_0$ which coincides with the ADM mass.


\subsection{Einstein-Born-Infeld theory}

We then consider EBI theory with the cosmological constant $\Lambda$
\be
\label{ebi}
G_2=-\beta\left(\sqrt{1-\frac{2{\cal F}}{\beta}}-1\right)-\frac{\Lambda}{8\pi G},
\qquad 
G_4=\frac{1}{16\pi G},
\qquad 
G_3=c_2=0,
\ee
where $\beta>0$ represents the nonlinear coupling parameter.
As the unique static and spherically symmetric BH solution in EM theory \eqref{ebi}, there exists the exact BH solution
\be
\label{ebi_sol}
h(r)
=
f(r)
&=&
\frac{1}{3}
\Big[
3
+
\left(8\pi G \beta-\Lambda \right)r^2
-
8\pi G\sqrt{\beta (Q^2+r^4\beta)}
-
16\pi Q\sqrt{\beta} {}_2F_1\left[\frac{1}{4},\frac{1}{2},\frac{5}{4},-\frac{r^4 \beta}{Q^2}\right]
\nonumber\\
&&
-
\frac{r_g}{r}
\left(
3
+
\left(8\pi G \beta-\Lambda \right)r_g^2
-
8\pi G\sqrt{\beta (Q^2+r_g^4\beta)}
-
16\pi Q\sqrt{\beta} {}_2F_1\left[\frac{1}{4},\frac{1}{2},\frac{5}{4},-\frac{r_g^4 \beta}{Q^2}\right]
\right)
\Big],
\nonumber\\
A_0(r)
&=&
-
\sqrt{\beta}
r
{}_2F_1\left[\frac{1}{4},\frac{1}{2},\frac{5}{4},-\frac{r^4 \beta}{Q^2}\right]
+q_{\rm EBI},
\quad 
A_1(r)
=
0,
\ee
where ${}_2F_1[\alpha,\beta,\gamma,x]$ represents the hypergeometric function
and the constant $q_{\rm EBI}$ corresponds to the gauge DOF.
In the limit of $\beta\to\infty$,
\be
f=h
&=&
-
\frac{\Lambda}{3}r^2
+
1
-
\frac{r_g}{r}
\left(
1
-
\frac{\Lambda r_g^2}{3}
+
\frac{4\pi GQ^2}{r_g^2}
\right)
+
\frac{4\pi GQ^2}{r^2}
+
{\cal O}
\left(
\frac{1}{\beta}
\right),
\nonumber\\
A_0(r)
&=&
q_{\rm EBI}'
+
\frac{Q}{r}
+
{\cal O}
\left(
\frac{1}{\beta}
\right),
\ee
which recovers the RN-(A)dS solution, where the constant $q_{\rm EBI}'$ corresponds to the redefined gauge DOF by taking the constant shift of $q_{\rm EBI}$ into account.
The variation of the Hamiltonian yields
\be
\label{var_horizon_3}
\delta{\cal H}_H
&=&
\frac{\delta r_g}{2G}
\left[
1
-
r_g^2\Lambda
+
8\pi  G\beta
\left(r_g^2
-
\sqrt{\frac{Q^2+r_g^4\beta}{\beta}}
\right)
\right]
+
4\pi  A_0(r_g) \delta Q.
\ee
However, the same term proportional to $\delta Q$ also appears in the variation of the Hamiltonian at the infinity $r\to\infty$, $\delta {\cal H}_\infty$. 
Thus, in the conservation of the Hamiltonian, the term proportional to $\delta Q$ cancels.
The differential of the BH entropy is given by 
\be
T_{\textsf{H}({\rm EBI})} 
\delta S_{\rm EBI}
=
\frac{\delta r_g}{2G}
\left[
1
-
r_g^2\Lambda
+
8\pi  G\beta
\left(r_g^2
-
\sqrt{\frac{Q^2+r_g^4\beta}{\beta}}
\right)
\right],
\ee
where the Hawking temperature \eqref{hawking_temperature} is given by 
\be
T_{\textsf{H}({\rm EBI})}
=
T_0
\left[
1
-
r_g^2\Lambda
+
8\pi  G\beta
\left(r_g^2
-
\sqrt{\frac{Q^2+r_g^4\beta}{\beta}}
\right)
\right].
\ee
Thus, we obtain the Bekenstein-Hawking entropy $S_{\rm EBI}=S_0$.
The thermodynamic mass of the BH is given by 
\be
M_{\rm EBI}
&=&
M_0
\left(
1
-
\frac{1}{3}
r_g^2\Lambda
+
\frac{8\pi G\left(\beta r_g^2-\sqrt{\beta (Q^2+r_g^4\beta)}\right)}
        {3} 
       {r_g}
-
\frac{16\pi GQ\sqrt{\beta}}{3} {}_2F_1\left[\frac{1}{4},\frac{1}{2},\frac{5}{4},-\frac{r_g^4 \beta}{Q^2}
\right]
\right),
\ee
which coincides with the ADM mass.
The conservation of the Hamiltonian $\delta {\cal H}=0$ yields the first law of thermodynamics for the electrically charged BH, $T_{\textsf{H}({\rm EBI})} \delta S_{\rm EBI}=\delta M_{\rm EBI}-\Phi_{H}\delta Q$, where $\Phi_H:=-4\pi \left(A_0(r\to \infty)-A_0(r_g)\right)$.

EBI theory \eqref{ebi} also admits the Schwarzschild-(A)dS solution with the trivial vector field \eqref{em_gr}.
We note that by applying the general stability criteria for electrically charged BH solutions in the presence of the nonlinear electrodynamics minimally coupled to the Einstein-Hilbert term derived in Ref.~\cite{Moreno:2002gg}, the linear stability of the BH solutions in EBI theory \eqref{ebi_sol} has been proven in Ref. \cite{Breton:2007bza} (see also Ref. \cite{Nomura:2020tpc}).

\section{Stealth Schwarzschild and Schwarzschild-(anti-) de Sitter solutions}
\label{sec4}

In the section, we focus on the class of GP theories \eqref{particular_theory} physically equivalent to the theory~\eqref{action_nmc}, which admits the stealth Schwarzschild and  Schwarzschild-(A)dS solutions~\cite{Chagoya:2016aar,Minamitsuji:2016ydr,Heisenberg:2017xda,Heisenberg:2017hwb,Minamitsuji:2017aan}.

\subsection{Solutions with the vanishing cosmological constant and Proca mass}
\label{sec4a}

In this subsection, we focus on the case of the vanishing Proca mass and cosmological constant $m=\Lambda=0$  in the theory \eqref{particular_theory} or \eqref{action_nmc}.

\subsubsection{Stealth Schwarzschild solution}
\label{sec4a1}

For arbitrary value of $\beta$, there exists the (electrically neutral) stealth Schwarzschild solution \cite{Chagoya:2016aar,Minamitsuji:2016ydr,Heisenberg:2017xda,Heisenberg:2017hwb} given by 
\be
\label{solution_stealth}
h(r)
=
f(r)
=
1
-
\frac{r_g}{r},
\qquad 
A_0(r)=q,
\qquad 
A_1(r)
=
q\frac{\sqrt{rr_g}}{r-r_g}
\Longrightarrow
Y(r)=\frac{q^2}{2}.
\ee
The variation of the Hamiltonian on the horizon $r=r_g$ yields
\be
\label{var_horizon_stealth}
\delta{\cal H}_H
=
\frac{1}{2G}
\left(1-8\pi Gq^2\beta\right)\delta r_g-16\pi \beta  r_g A_0(r_g)\delta q.
\ee
We emphasize that following the formulation in Ref. \cite{Minamitsuji:2023nvh}, the same differential as Eq. \eqref{var_horizon_stealth} can be obtained from the stealth BH solution in shift-symmetric Horndeski theory with $G_2=2m^2 X-\frac{\Lambda}{8\pi G}$, $G_4=\frac{1}{16\pi G}+\beta X$ and $G_3=G_5=0$ for the linearly time-dependent ansatz of the scalar field $\varphi=qt+\psi (r)$.
Since the same $\delta q$ term as that in Eq. \eqref{var_horizon_stealth} also appears in the variation of the Hamiltonian at the infinity $r=\infty$, these terms cancel in the conservation of the Hamiltonian $\delta {\cal H}=0$.
Thus, the differential of the BH entropy is given by 
\be
\label{var_horizon_em_g}
T_{\textsf{H}({\rm GP})} \delta S_{\rm GP}
=
\frac{1}{2G}
\left(1-8\pi Gq^2\beta\right)\delta r_g,
\ee
where the Hawking temperature \eqref{hawking_temperature} is given by that of the Schwarzschild spacetime
$T_{\textsf{H}({\rm GP})}=T_0$.
Thus, we obtain the entropy of the stealth Schwarzschild BH 
\be
S_{\rm GP}
=S_0\left(1-8\pi Gq^2\beta\right).
\ee
The variation of the thermodynamics mass $\delta M_{\rm GP}$ is given by $\delta M_{\rm GP}=\frac{1}{2G} \left(1-8\pi Gq^2\beta\right)\delta r_g$.
Therefore, the thermodynamic mass of the BH is given by 
\be
M_{\rm GP}
=
M_0
\left(
1-8\pi G q^2\beta 
\right).
\ee
We note that $M_0$ corresponds to the ADM mass. For the positivity of the entropy and thermodynamic mass of the BH, we have to impose $8\pi G q^2\beta<1$.

There also exists the GR Schwarzschild solution $h(r)=f(r)=1-\frac{r_g}{r}$ with the trivial vector field $A_\mu=0$.
The entropy of the GR Schwarzschild BH is given by the Bekenstein-Hawking entropy
\be
S_{\rm GR}=S_0,
\ee
and the thermodynamic mass of the BH is given by 
\be
\label{bcl_mass_gr}
M_{\rm GR}
=
M_0.
\ee
When we compare these two entropies ($S_{\rm GR}\,,S_{\rm GP}$) at the same thermodynamic mass of BHs $M_{\rm GR}=M_{\rm GP}$, we can easily find 
\begin{eqnarray}
S_{\rm GP}
=
\frac{S_{\rm GR}}
    {1-8\pi G q^2\beta}.
\end{eqnarray}
Thus, the stealth Schwarzschild solution in GP theories is thermodynamically more stable than the GR Schwarzschild BH when $\beta>0$.

\subsubsection{
Stealth Schwarzschild solution with the electric field}
\label{sec4a2}

For the specific value $\beta=\frac{1}{4}$, there also exists the stealth Schwarzschild solution with the electric field \cite{Chagoya:2016aar,Minamitsuji:2016ydr,Heisenberg:2017xda,Heisenberg:2017hwb} given by 
\be
\label{solution_stealth2}
h(r)
=
f(r)
=
1
-
\frac{r_g}{r},
\qquad 
A_0(r)=q+\frac{Q}{r},
\qquad 
A_1(r)
=
\frac{\sqrt{Q^2+2qQ r+q^2rr_g}}{r-r_g}
\Longrightarrow
Y(r)=\frac{q^2}{2}.
\ee
We note that although a BH has nonzero electric charge $Q$, the metric remains the Schwarzschild one unlike the case of EM theory.
The variation of the Hamiltonian on the horizon $r=r_g$ yields \footnote{This result differs from Ref.~(48) in Eq. \cite{Chagoya:2023ddb} after the replacement of $q\to P$ and $r_g\to 2m$.}
\be
\label{var_horizon_charged_stealth}
\delta{\cal H}_H
=
\frac{1}{2G}
\left(
1-2\pi Gq^2
\right)
\delta r_g
-
4\pi
r_g A_0(r_g)
\delta q.
\ee
Since the same $\delta q$ term as that in Eq. \eqref{var_horizon_charged_stealth} also appears in the variation of the Hamiltonian at the infinity $r=\infty$, these terms cancel in the conservation of the total Hamiltonian $\delta {\cal H}=0$.
Thus, the differential of the BH entropy is solely given by 
\be
\label{var_horizon_em2}
T_{\textsf{H}({\rm GP})} 
\delta S_{\rm GP}
=
\frac{1}{2G}
\left(
1-2\pi Gq^2
\right)
\delta r_g.
\ee
Since the Hawking temperature is given by Eq.~\eqref{T0}, we obtain the entropy of the  
stealth Schwarzschild BH with the electric field as
\be
S_{\rm GP}=S_0\left(1-2\pi Gq^2\right).
\ee
On the other hand, by integrating $\delta M_{\rm GP}=\frac{1}{2G}\left(1-2\pi Gq^2\right)\delta r_g$, the thermodynamic mass of the BH is given by 
\be
M_{\rm GP}=M_0\left(1-2\pi q^2 G\right).
\ee
Thus, despite the presence of the electric charge $Q$, the differentials of the entropy and the thermodynamic mass of the BH are integrable with respect to $r_g$ and these thermodynamic quantities exactly coincide with the $\beta=\frac{1}{4}$ limit of the electrically neutral case discussed in Sec. \ref{sec4a1}.
In other words, the electric charge $Q$ does not affect thermodynamics of BHs as well as the background metric.
For the positivity of the entropy and the thermodynamic mass of the BH, we have to impose $0<2\pi G q^2<1$.

When we compare these two entropies ($S_{\rm GR}\,,S_{\rm GP}$) at the same mass value $M_{\rm GR}=M_{\rm GP}$, we can easily find that $S_{\rm GP}=\frac{S_{\rm GR}} {1-2\pi G q^2}$ for $0<2\pi G q^2<1$.
Hence, $S_{\rm GP}>S_{\rm GR}$ and the 
stealth Schwarzschild BH with the electric field is thermodynamically more stable than the GR Schwarzschild BH.

\subsubsection{Solution with $A_1(r)=0$}
\label{sec4a3}

For $\beta=\frac{1}{4}$, there also exists the exact BH solution in the branch with $A_1(r)=0$ \cite{Geng:2015kvs,Minamitsuji:2016ydr} given by 
\be
\label{sqrt_bh}
f(r)=h(r)
=
1-\sqrt{\frac{r_g}{r}},
\qquad
A_0(r)=\frac{1}{\sqrt{2\pi G}} f(r),
\qquad 
A_1(r)
=
0.
\ee
Because of the absence of the $1/r$ term in the metric functions, the ADM mass for the solution \eqref{sqrt_bh} $M_{\rm ADM}=\frac{1}{2G} \lim_{r\to \infty} r\left(1-f(r)\right)$ diverges.
Thus, the metric of the solution \eqref{sqrt_bh} is not asymptotically flat.

The variation of the Hamiltonian on the horizon $r=r_g$ yields
\be
\label{var_horizon_em_g2}
\delta{\cal H}_H
=
T_{\textsf{H}({\rm GP})} \delta S_{\rm GP}
=
\frac{1}{4G}\delta r_g,
\ee
where the Hawking temperature \eqref{hawking_temperature} is given by that of the Schwarzschild spacetime
$T_{\textsf{H}({\rm GP})}=\frac{1}{2}T_0$.
Thus, we obtain the Bekenstein-Hawking formula for the BH entropy $S=S_0$.
The differential of the thermodynamic mass of the BH is given by $\delta{\cal H}_\infty=\delta M_{\rm GP}=\frac{1}{4G} \delta r_g$.
Therefore, the thermodynamic mass of the BH is given by $M_{\rm GP}=\frac{M_0}{2}$.
While the ADM mass for the solution \eqref{sqrt_bh} diverges, the thermodynamic mass of BHs $M_{\rm GP}$ remains finite.
Our results coincide with the thermodynamic properties obtained in Ref. \cite{Geng:2015kvs}.

We now compare the entropy of the BH solution \eqref{sqrt_bh} with that of the GR Schwarzschild solution with the trivial vector field.
For the same thermodynamic mass of BHs $M_{\rm GP}=M_{\rm GR}$, we find that $S_{\rm GP}=4S_{\rm GR}$ and hence the entropy of the BH solution \eqref{sqrt_bh} is always larger than that of the GR Schwarzschild solution.
On the other hand, if we compare the BH solution \eqref{sqrt_bh} with the 
stealth Schwarzschild solution with the electric field \eqref{solution_stealth2} discussed in Sec. \ref{sec4a2} , the entropy of the BH solution \eqref{sqrt_bh} is larger than that of the stealth Schwarzschild solution 
with the electric field \eqref{solution_stealth2} for $0<q^2<\frac{3}{8\pi G}$, while smaller for the opposite case $\frac{3}{8\pi G}<q^2<\frac{1}{2\pi G}$.

As the consequence of a naive comparison of entropies for the three different BH solutions existing in the theory \eqref{particular_theory} (or equivalently \eqref{action_nmc}) with $m=\Lambda=0$ and $\beta=\frac{1}{4}$, the solution \eqref{sqrt_bh} seems to be thermodynamically preferred.
However, as mentioned above, the solution \eqref{sqrt_bh} is not asymptotically flat, because in the large distance regions the leading corrections to the metric functions are given by $\frac{1}{\sqrt{r}}$, not by $\frac{1}{r}$.
Thus, the solution \eqref{sqrt_bh} does not share the same asymptotic behaviour of the spacetime with the other BH solutions in the same theory.
Since a phase transition could hardly affect the asymptotic structure of the spacetime, we expect that the transition between the BH solution of Eq.~\eqref{sqrt_bh} and the 
stealth Schwarzschild BH with the electric field \eqref{solution_stealth2} or the GR Schwarzschild BH solution would not be plausible.

\subsection{Solutions with the nonzero cosmological constant and Proca mass }

In this subsection, in the theory \eqref{particular_theory} or equivalently \eqref{action_nmc}, we focus on the solutions in the case of $m\neq 0$ and $\Lambda\neq 0$, which allow asymptotically dS or AdS BH solutions.
Besides the BH solutions with the nontrivial vector field, there are the GR Schwarzschild-(A)dS solutions with the trivial vector field given by 
\be
\label{solution_GRSdS}
h(r)
&=&
f(r)
=
-
\frac{{ \Lambda}}{3}r^2
+1
-
\frac{r_g}{r}
\left(
1
-
\frac{{\Lambda}}{3}r_g^2
\right),
\quad 
A_0(r)
=
A_1(r)
=
0.
\ee
The entropy and the thermodynamic mass of the GR Schwarzschild-dS BH are given by 
\be
\label{ent_GRSdS}
S_{\rm GR}
=
S_0,
\qquad
M_{\rm GR}=M_0\left(1-\frac{\Lambda r_g^2}{3}\right),
\ee
where $S_0$ and $M_0$ are given by Eqs. \eqref{S0} and \eqref{M0}, respectively.

\subsubsection{Schwarzschild-(A)dS solution}
\label{sec4b1}

For arbitrary value of $\beta$, there exists the Schwarzschild-(A)dS BH solution \cite{Minamitsuji:2016ydr,Heisenberg:2017xda,Heisenberg:2017hwb} given by 
\be
\label{solution4}
h(r)
&=&
f(r)
=
-
\frac{{\bar \Lambda}}{3}r^2
+1
-
\frac{r_g}{r}
\left(
1
-
\frac{{\bar \Lambda}}{3}r_g^2
\right),
\nonumber
\\
A_0(r)&=&
\frac{\sqrt{\Lambda-{\bar \Lambda}}}{4m\sqrt{\pi G}},
\quad 
A_1(r)
=
\frac{\sqrt{\Lambda-{\bar \Lambda}}}{4m\sqrt{\pi G}}
\frac{\sqrt{1-f(r)}}
       {f(r)}
\Longrightarrow
Y(r)=
\frac{\Lambda-{\bar \Lambda}}{32\pi G m^2},
\ee
where the effective cosmological constant is given by 
\be{\bar \Lambda}:=-\frac{m^2}{\beta}, 
\label{barLambda}
\ee
and the existence of the solution requires $\Lambda-{\bar \Lambda}\geq 0$.
The variation of the Hamiltonian on the horizon $r=r_g$ yields
\be
\label{var_horizon_em3}
\delta{\cal H}_H
=
T_{\textsf{H}({\rm GP})} \delta S_{\rm GP}
=
\frac{\delta r_g}{4G \beta}
\left(
m^2r_g^2+\beta
\right)
\left(
1
+
\frac{\Lambda}{\bar \Lambda}
\right),
\ee
where the Hawking temperature \eqref{hawking_temperature} is given by $T_{\textsf{H}({\rm GP})}=T_0\left(1-{\bar \Lambda}r_g^2\right)$.
By integrating Eq. \eqref{var_horizon_em3}, we obtain the BH entropy 
\be
S_{\rm GP}
=
S_0
\frac{\Lambda+{\bar \Lambda}}{2{\bar \Lambda}}.
\ee
The variation of the Hamiltonian at the infinity $r\to \infty$ yields
\be
\label{var_infinity_em2}
\delta {\cal H}_\infty
=
\delta M_{\rm GP}
=\frac{\delta r_g}{4G}
\left(
1+\frac{m^2r_g^2}{\beta}
\right)
\frac{\Lambda+{\bar \Lambda}}
        {{\bar \Lambda}},
\ee
and by integrating this the thermodynamic mass of the BH is given by 
\be
M_{\rm GP}
=
M_0
\left(
1-\frac{\bar \Lambda}{3}r^2
\right)
\frac{\Lambda+{\bar \Lambda}}  {2{\bar \Lambda}}.
\ee
Thus, we have the two different Schwarzschild-(A)dS solutions with the thermodynamical properties as summarized in Table \ref{table1}.
\begin{table}[htbp]
\begin{center}
\scalebox{1.1}[1.1]{
  \begin{tabular}{|c|c|c|c|}
\hline 
solution
& mass
&
entropy
&
temperature
\\
\hline 
Schwarzschild-(A)dS solution with $\Lambda$ (GR)
&
$\left(1-\frac{\Lambda r_g^2}{3}\right)M_0$
&
$S_0$
&
$\left(1-\Lambda r_g^2\right)T_{0}$
\\
\hline 
Schwarzschild-(A)dS solution with ${\bar \Lambda}$ (GP)
&
$\left(1-\frac{{\bar \Lambda} r_g^2}{3}\right)\frac{\Lambda+{\bar\Lambda}} {2\bar\Lambda}M_0 $
&
$\frac{\Lambda+\bar\Lambda}     {2\bar\Lambda}S_0$
&
$\left(1-\bar \Lambda r_g^2\right)T_0$
\\
\hline
 \end{tabular}
 }  
 \caption{Summary of thermodynamic quantities of Schwarzschild-(A)dS solutions.}
\label{table1}
\end{center}
\end{table}

Since the metric solutions and the thermodynamic quantities associated with them remain the same as those in the counterpart solutions in shift-symmetric Horndeski theories, our discussion below will have a considerable overlap with that in Sec. V of Ref. \cite{Minamitsuji:2023nvh}.
In this paper, we summarize only the essential properties of them.

\begin{enumerate}

\item{\texorpdfstring{In the case of $\Lambda\ge \bar\Lambda>0$ for $\beta<0$}{TEXT}}

For a given mass $M$, the entropy of the GR Schwarzschild-dS BH with $\Lambda$ is always larger than that of GP Schwarzschild-dS BH  with $\bar \Lambda$.
This means that GP Schwarzschild-dS BH is more thermodynamically unstable than the GR Schwarzschild-dS BH. 
For the comparison of these two entropies, readers should refer to Fig 1 of Ref. \cite{Minamitsuji:2023nvh}.

\item{\texorpdfstring{In the case of ${\bar\Lambda}\le \Lambda<0$ for $\beta>0$:}{TEXT}}

In this case, $S_{\rm GR}(M)$ and $S_{\rm GP}(M)$ coincide at some critical mass $M_{\rm GR\mathchar`-GP}$, beyond which $S_{\rm GR}>S_{\rm GP}$. 
In an asymptotically AdS spacetime, there exists another critical mass $M_{\textsf{HP}}$, below which the Schwarzschild-AdS BH evaporates to thermal radiation in AdS space via the Hawking-Page (HP) transition~\cite{Hawking:1982dh}. 
In the present case, we find two critical masses, $M_{\textsf{HP} ({\rm GR})}$ and $M_{\textsf{HP} ({\rm GP})}$ which satisfy $M_{\textsf{HP} ({\rm GR})}>M_{\textsf{HP} ({\rm GP})}$. 
We also find $M_{\rm GR\mathchar`-GP}<M_{\textsf{HP} ({\rm GR})}$.
As a result, we can classify into two cases: (1) $M_{\textsf{HP} ({\rm GR})}>M_{\textsf{HP} ({\rm GP})}>M_{\rm GR\mathchar`-GP}$, and  (2) $M_{\textsf{HP} ({\rm GR})}>M_{\rm GR\mathchar`-GP}>M_{\textsf{HP} ({\rm GP})}$.
For convenience, we introduce the AdS curvature radii $\ell:=\sqrt{-3/\Lambda}$ and $\bar \ell :=\sqrt{-3/\bar \Lambda}$.

If $\mathfrak{r}_{\rm cr}<\bar \ell/\ell <1$, we find case (1) below, while if $0<\bar \ell/\ell <\mathfrak{r}_{\rm cr}$, we obtain case (2) below, where $\mathfrak{r}_{\rm cr}$ is given by the root of the equation $\mathfrak{r}_{\rm cr}^6 + 3 \mathfrak{r}_{\rm cr}^4+ 16 \mathfrak{r}_{\rm cr}^2-4  = 0$, i.e., $\mathfrak{r}_{\rm cr}\approx 0.48835$.
\begin{itemize}
\item Case (1), $\mathfrak{r}_{\rm cr}<\bar \ell/\ell <1$:
\\
For $M<M_{\rm GR\mathchar`-GP}$, we find only thermal radiation in AdS with the effective cosmological constant $\bar \Lambda$, 
For $M_{\rm GR\mathchar`-GP}<M<M_{\textsf{HP} ({\rm GP})}$, it is also thermal radiation in AdS space with the cosmological constant $\Lambda$. 
For $M_{\textsf{HP} ({\rm GR})}>M>M_{\textsf{HP} ({\rm GP})}$, GP Schwarzschild-AdS BH will evaporate via the HP phase transition, finding thermal radiation in AdS space with the cosmological constant $\Lambda$. 
For $M>M_{\textsf{HP} ({\rm GR})}$, GP Schwarzschild-AdS BH will evolve into the GR Schwarzschild-AdS BH via thermal phase transition.

\item Case (2),  $0<\bar \ell/\ell <\mathfrak{r}_{\rm cr}$:
\\
For $M<M_{\textsf{HP} ({\rm GP})}$, there is only thermal radiation in AdS space with the effective cosmological constant $\bar \Lambda$ just as in Case (1). 
For $M_{\textsf{HP} ({\rm GP})}<M<M_{\rm GR\mathchar`-GP}$, we find the transition from thermal radiation in AdS space with $\Lambda$ into the stable GP Schwarzschild-AdS BH. 
For $M_{\textsf{HP} ({\rm GR})}>M>M_{\rm GR\mathchar`-GP}$,  GP Schwarzschild-AdS BH will evaporate into thermal radiation in AdS space with $\Lambda$.
For $M>M_{\textsf{HP} ({\rm GR})}$, GP Schwarzschild-AdS BH will evolve into the GR Schwarzschild-AdS BH via thermal phase transition just as in Case (1).
\end{itemize}
For the more concrete comparison of two entropies in each case, readers should refer to Fig 2 of Ref. \cite{Minamitsuji:2023nvh}.

\end{enumerate}

\subsubsection{Charged Schwarzschild-AdS solution}
\label{sec4b2}

For $\beta=\frac{1}{4}$, there exists the charged Schwarzschild-AdS BH solution \cite{Minamitsuji:2016ydr,Heisenberg:2017xda,Heisenberg:2017hwb},
given by 
\be
\label{solution42}
h(r)
&=&
f(r)
=
-
\frac{\bar \Lambda}{3}r^2
+1
-
\frac{r_g}{r}
\left(
1
-
\frac{\bar \Lambda}{3}r_g^2
\right),
\nonumber
\\
A_0(r)&=&
\frac{Q}{r}
+
\frac{1}{4m\sqrt{\pi}}
\sqrt{
\frac{
\Lambda-{\bar\Lambda}
}{G}
},
\quad 
A_1(r)
=
\frac{1}{\sqrt{f(r)}}
\sqrt{
\frac{A_0(r)^2}{f(r)}
-
\frac{
\Lambda-{\bar\Lambda}}
{16\pi G m^2}}
\Longrightarrow
Y(r)=
\frac{\Lambda-{\bar \Lambda}}{32\pi G m^2},
\ee
where the effective cosmological constant is given by 
\be
\label{barLambda2}
{\bar \Lambda}:=-4m^2, 
\ee
$Q$ is the electric charge, and
the existence of the solution requires $\Lambda-{\bar \Lambda}\geq 0$.
The variation of the Hamiltonian on the horizon $r=r_g$ yields
\be
\label{var_horizon_em4}
\delta{\cal H}_H
=
T_{\textsf{H}({\rm GP})} 
\delta S_{\rm GP}
=
-
\frac{\delta r_g}
        {16G m^2}
\left(
1+4m^2r_g^2
\right)
\left(
\Lambda+{\bar \Lambda}
\right),
\ee
where the Hawking temperature \eqref{hawking_temperature} is given by $T_{\textsf{H}({\rm GP})}=T_0\left(1+4m^2r_g^2\right)$.
Thus, we obtain the integrable relation 
\be
\delta S_{\rm GP}
=
\frac{\pi r_g\delta r_g}{G}
\frac{\Lambda+{\bar \Lambda}}
       {{\bar\Lambda}}
\Longrightarrow
S_{\rm GP}
=
\frac{S_0}{2}
\frac{\Lambda+{\bar \Lambda}}{{\bar\Lambda}}.
\ee
The variation of the Hamiltonian at the infinity $r\to \infty$ yields
\be
\label{var_infinity_em3}
\delta {\cal H}_\infty
=
\delta M_{\rm GP}
=\frac{\delta r_g}{4G}
\left(
1+4 m^2r_g^2
\right)
\frac{\Lambda+{\bar \Lambda}}
        {{\bar \Lambda}}.
\ee
The mass of the total system is given by 
\be
M_{\rm GP}
=
\frac{M_0}{2}
\left(
1-\frac{{\bar \Lambda}r_g^2}{3}
\right)
\frac{\Lambda+{\bar \Lambda}}
        {{\bar \Lambda}}.
\ee
Thus, despite the presence of the nonzero electric charge $Q$, the differentials of the entropy and mass are integrable with respect to $r_g$ and the results correspond to the $\beta=\frac{1}{4}$ limit of the electrically neutral case.
The thermodynamical stability of the Schwarzschild-AdS BH solutions remains the same as that discussed in Sec.~\ref{sec4a2}.

\section{Other black hole solutions in GP theories}
\label{sec5}

\subsection{Schwarzschild-de Sitter solution with generalized cubic-order GP interaction}
\label{sec5a}

We consider the class of GP theories with generalized cubic-order GP interaction
\be
\label{generalized_cubic_model}
G_2=g_2(Y),
\qquad 
G_3=g_3(Y),
\qquad 
G_4=\frac{1}{16\pi G},
\qquad 
c_2=0,
\ee
where $g_2(Y)$ and $g_3(Y)$ are the regular functions of $Y$.
The theory \eqref{generalized_cubic_model} admits the exact Schwarzschild-(A)dS solution \cite{Cisterna:2016nwq,Minamitsuji:2017aan}
\be
\label{generalized_cubic_sol}
h(r)
&=&
f(r)
=
\frac{8\pi G g_2(Y_0)}{3}r^2
+1
-
\frac{r_g}{r}
\left(
1
+
\frac{8\pi G r_g^2 g_2(Y_0)}{3}
\right),
\nonumber\\
A_0(r)
&=&
\sqrt{
\frac{1}{9}
\left(
48\pi G Y_0 g_2(Y_0)+\frac{g_{2Y}(Y_0)^2}{g_{3Y}(Y_0)^2}
\right)r^2
+
2Y_0
+
\frac{1}{r}
\left(
-\frac{2r_g Y_0}{3}
\left(
3+8\pi G r_g^2 g_2(Y_0)
\right)
-
\frac{2P g_{2Y}(Y_0)}
        {3g_{3Y}(Y_0)}
\right)
+
\frac{P^2}{r^4}
},
\nonumber\\
A_1(r)
&=&
\frac{1}{r^2f(r)}
\left(
P
-
\frac{r^3}{3}
\frac{g_{2Y}(Y_0)}
        {g_{3Y}(Y_0)}
\right),
\ee
where 
$P$ and $Y_0$ are integration constants. 
The constant norm of the vector field $Y=Y_0$ is obtained as the unique solution of the equations of motion, and the solution \eqref{generalized_cubic_sol} is the unique one for the BH with nontrivial profile of the vector field.
The sign of $Y_0$, i.e., the character of the vector field, cannot be determined by the equations of motion.
The cubic-order coupling function $g_3(Y)$ does not contribute to the spacetime geometry, while the asymptotic behavior of the spacetime is fixed by the value of $g_2(Y_0)$.
We note that the constant $P$ may be interpreted as the electric charge which does not affect the metric functions $h(r)$ and $f(r)$, and hence the solution \eqref{generalized_cubic_sol} may be interpreted as a kind of stealth BHs.

The variation of the Hamiltonian on the horizon $r=r_g$ yields
\be
\label{var_horizon_3_2}
\delta{\cal H}_H
=
T_{\textsf{H}({\rm GP})} 
\delta S_{\rm GP}
=
\frac{\delta r_g}{2G}
\left(
1
+
8\pi G r_g^2g_2(Y_0)
\right),
\ee
where the Hawking temperature \eqref{hawking_temperature} is given by 
\be
T_{\textsf{H}({\rm GP})}
=
T_0
\left(
1+8\pi G g_2(Y_0)r_g^2
\right).
\ee
In Eq. \eqref{var_horizon_3} the variation with respect to $P$ cancels and does not contribute to the BH entropy.
The variation of $P$ also does not contribute to the thermodynamic mass of BHs.
Thus, by integrating this, we obtain the Bekenstein-Hawking entropy as the entropy of the BH solution \eqref{generalized_cubic_sol} $S_{\rm GP}=S_0$.

The variation of the Hamiltonian at the infinity $r\to \infty$ yields
\be
\label{var_infinity_em4}
\delta {\cal H}_\infty
=
\delta  M_{\rm GP}
=
\frac{\delta r_g}{2}
\left(
\frac{1}{G}
+
8\pi r_g^2g_2(Y_0)
\right).
\ee
The thermodynamic mass of the BH is given by 
\be
M_{\rm GP}
=
M_0
\left(1+\frac{8\pi G}{3} r_g^2 g_2(Y_0)\right),
\ee
which coincide with the ADM mass $M_{\rm ADM}=M_{\rm GP}$.

From now on, we focus on the specific choice of the coupling functions
\be
\label{specifc_cubic_model}
g_2(Y)=-\frac{\Lambda}{8\pi G}+ 2m^2 Y,
\qquad 
g_3(Y)=\gamma_3 Y,
\ee
where $m^2$ and $\gamma_3$ represent the mass term of the vector field and the coupling constant of the cubic-order interaction, the entropy and the thermodynamic mass of BHs, respectively, reduce to 
\be
S_{\rm GP}
=
S_0,
\qquad
M_{\rm GP}
=
\left(
1
+
\frac{r_g^2}{3}
\left(
16\pi G Y_0m^2
-
\Lambda
\right)
\right)
M_0.
\ee
The theory \eqref{specifc_cubic_model} also admits the GR BH solution with the trivial vector field \eqref{solution_GRSdS}
whose entropy and thermodynamic mass are, respectively, given by Eq.~\eqref{ent_GRSdS}.
For $\Lambda>0$, when $M_{\rm GR}=\frac{1}{3G\sqrt{\Lambda}}$, the condition $f(r)=h(r)=0$ has the degenerate solution, i.e., the BH and cosmological horizons coincide at $r_g=\frac{1}{\sqrt{\Lambda}}$.
Thus, for the horizons to be formed for the GR Schwarzschild-dS solution, we restrict
\be
\label{bound_on_rg}
0<r_{g}<\frac{1}{\sqrt{\Lambda}}.
\ee
On the other hand, for $\Lambda<0$, the event horizon is always present in the GR Schwarzschild-AdS solution and there is no upper bound on the value of $r_g$.
However, since heat capacities for the Schwarzschild-AdS BHs in the two theories are, respectively, given by 
\be
C_{\rm GP}
:=\frac{\partial M_{\rm GP}}{\partial T_{\textsf{H}({\rm GP})}}
=
\frac{2\pi r_g^2}{G}
\frac{1+r_g^2\left(16\pi G m^2 Y_0+|\Lambda|\right)}
        {-1+r_g^2\left(16\pi G m^2 Y_0+|\Lambda|\right)},
\qquad 
C_{\rm GR}
:=\frac{\partial M_{\rm GR}}{\partial T_{\textsf{H}({\rm GR})}}
=
\frac{2\pi r_g^2}{G}
\frac{1+r_g^2|\Lambda|}
        {-1+r_g^2|\Lambda|}.
\ee
Thus, we find that $C_{\rm GP}<0$ and $C_{\rm GR}<0$ for
\be
r_g<\sqrt{\frac{1}
                      {|\Lambda|+16\pi G m^2Y_0}},
\qquad
r_g<\sqrt{\frac{1}{|\Lambda|}},
\ee
respectively, where an AdS BH decays into thermal radiation through the HP phase transition \cite{Hawking:1982dh}.

For the same thermodynamic mass of BHs, $M_{\rm GP}=M_{\rm GR}$, we compare the entropies  for the solutions \eqref{generalized_cubic_sol} and \eqref{solution_GRSdS}.
For $\Lambda>0$, from Eq.~\eqref{bound_on_rg}, we impose  $0<r_{g({\rm GR})}<\frac{1}{\sqrt{\Lambda}}$.
The two horizon radii $r_{g({\rm GP})}$ and $r_{g({\rm GR})}$ are related by
\be
\label{horizon_radii}
r_{g({\rm GP})}
\left(
1
-
\frac{r_{g({\rm GP} )}^2}{3}
\left(
\Lambda
-
16\pi G Y_0m^2
\right)
\right)
=
r_{g({\rm GR})}
\left(
1
-
\frac{r_{g({\rm GR} )}^2}{3}
\Lambda
\right).
\ee
It is easy to check that, for the spacelike vector field $Y_0<0$, $r_{g({\rm GP} )}>r_{g({\rm GR} )}$ and hence $S_{\rm GP}>S_{\rm GR}$.
On the other hand,  for the timelike vector field $Y_0>0$, $S_{\rm GP}<S_{\rm GR}$.

\subsubsection{Schwarzschild-dS solutions}

In the case of the Schwazschild-dS solutions ($\Lambda>0$), we introduce the dimensionless horizon radii $x_{\rm GP}:=\sqrt{\frac{\Lambda}{3}} r_{g({\rm GP})}$
and $x_{\rm GR}:=\sqrt{\frac{\Lambda}{3}} r_{g({\rm GR})}$,
and the dimensional parameter $\Gamma_2:= \frac{16 \pi G Y_0m^2}{\Lambda}$, Eq.~\eqref{horizon_radii} reduces to 
\be
\label{gpgr}
x_{\rm GP}
\left(
1
-
x_{\rm GP}^2
\left(
1-
\Gamma_2
\right)
\right)
=
x_{\rm GR}
\left(
1
-
x_{\rm GR}^2\right).
\ee
For Eq.~\eqref{bound_on_rg}, we have to impose $0<x_{\rm GR}<\frac{1}{\sqrt{3}}$.
In order to obtain the Schwarzschild-dS solution in GP theory, we also impose that $\Gamma_2<1$.

\begin{figure}[tb]
  \begin{center}
      \includegraphics[keepaspectratio=true,height=50mm]{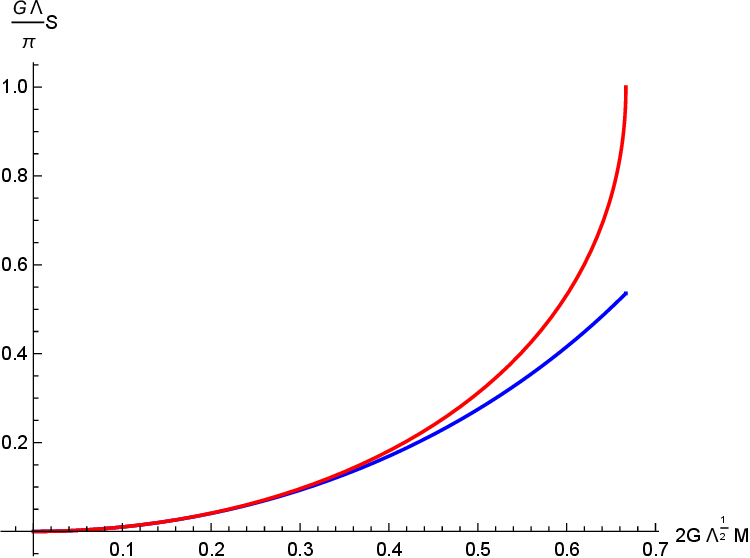}
      \includegraphics[keepaspectratio=true,height=50mm]{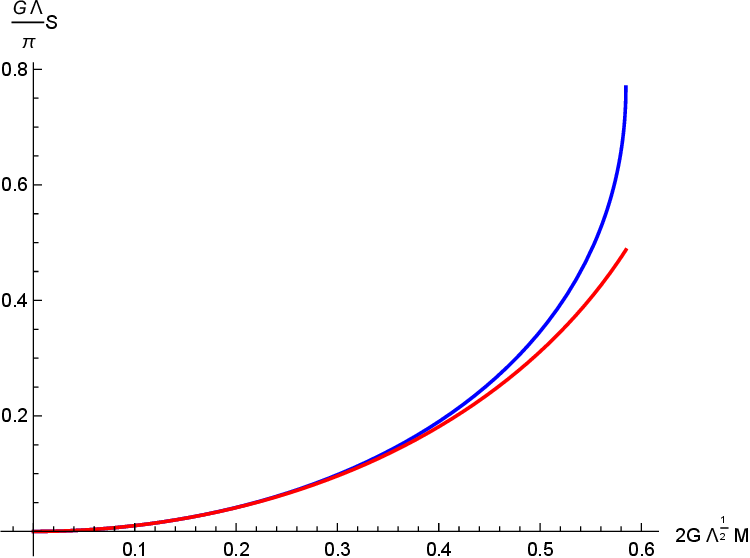}
      \caption{The normalized entropy of BHs $\frac{G\Lambda}{\pi} S$ is shown as the function of the normalized thermodynamic mass of BHs $2G\sqrt{\Lambda}M:=2G\sqrt{\Lambda}M_{\rm GP}=2G\sqrt{\Lambda}M_{\rm GR}$ for ${\Gamma}_2=0.50$ (in the left panel) and  ${\Gamma}_2=-0.30$ (in the right panel). In each panel, blue and red curves correspond to the normalized entropies for the solutions \eqref{generalized_cubic_sol} and \eqref{solution_GRSdS}, respectively.}
  \label{entropy_p}
  \end{center}
\end{figure}

In Fig. \ref{entropy_p}, the normalized entropy of BHs, $\frac{G\Lambda}{\pi} S$, is shown as the function of the normalized thermodynamic mass of BHs, $2G\sqrt{\Lambda}M:=2G\sqrt{\Lambda}M_{\rm GP}=2G\sqrt{\Lambda}M_{\rm GR}$ for ${\Gamma}_2=0.50$ (in the left panel) and  ${\Gamma}_2=-0.30$ (in the right panel). 
In each panel, red and blue curves correspond to the normalized entropies for the solutions \eqref{generalized_cubic_sol} and \eqref{solution_GRSdS}, respectively.
The ratio $S_{\rm GP}/S_{\rm GR}$ decreases as $Y_0$ increases, while increases as
$Y_0$ decreases.

\subsubsection{Schwarzschild-AdS solutions}

\begin{figure}[tb]
  \begin{center}
      \includegraphics[keepaspectratio=true,height=50mm]{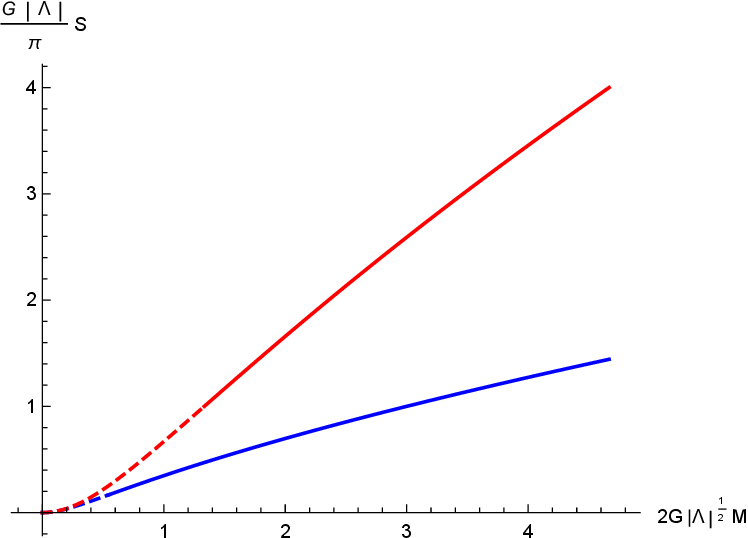}
      \includegraphics[keepaspectratio=true,height=50mm]{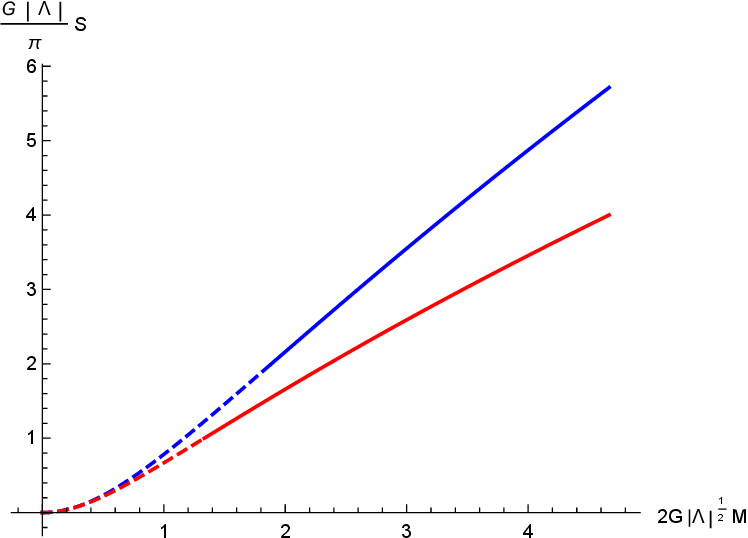}
      \caption{The normalized entropy of BHs, $\frac{G|\Lambda|}{\pi} S$, is shown as the function of the normalized thermodynamic mass of BHs, $2G\sqrt{|\Lambda|}M:=2G\sqrt{|\Lambda|}M_{\rm GP}=2G\sqrt{|\Lambda|}M_{\rm GR}$, for ${\tilde \Gamma}_2=5.0$ (in the left panel) and  ${\tilde \Gamma}_2=-0.50$ (in the right panel). 
In each panel, blue and red curves correspond to the normalized entropies for the solutions \eqref{generalized_cubic_sol} and \eqref{solution_GRSdS}, respectively.
For each curve, the solid and dashed regions correspond to those which are stable and unstable against the HP phase transition, respectively
}
  \label{entropy_m}
  \end{center}
\end{figure}

On the other hand, for the Schwarzschild-AdS solutions ($\Lambda<0$), we introduce the dimensionless horizon radii ${\tilde x}_{\rm GP}:=\sqrt{\frac{|\Lambda|}{3}} r_{g({\rm GP})}$ and ${\tilde x}_{\rm GR}:=\sqrt{\frac{|\Lambda|}{3}}r_{g({\rm GR})}$,
and the dimensional parameter ${\tilde \Gamma}_2:= \frac{16 \pi G Y_0m^2}{|\Lambda|}$, Eq.~\eqref{horizon_radii} reduces to 
\be
{\tilde x}_{\rm GP}
\left(
1
+
{\tilde x}_{\rm GP}^2
\left(
1+
{\tilde  \Gamma}_2
\right)
\right)
=
{\tilde x}_{\rm GR}
\left(
1
+
{\tilde x}_{\rm GR}^2
\right).
\ee
In order to have the Schwarzschild-AdS solutions, we have to impose that ${\tilde \Gamma_2}>-1$.
In Fig. \ref{entropy_m}, the normalized entropy of BHs, $\frac{G|\Lambda|}{\pi} S$, is shown as the function of the normalized thermodynamic mass of BHs, $2G\sqrt{|\Lambda|}M:=2G\sqrt{|\Lambda|}M_{\rm GP}=2G\sqrt{|\Lambda|}M_{\rm GR}$, for ${\tilde \Gamma}_2=5.0$ (in the left panel) and  ${\tilde \Gamma}_2=-0.50$ (in the right panel). 
In each panel, blue and red curves correspond to the normalized entropies for the solutions \eqref{generalized_cubic_sol} and \eqref{solution_GRSdS}, respectively.
For each curve, the solid and dashed regions correspond to those which are stable and unstable against the HP phase transition, respectively.
For $Y_0>0$, for a smaller thermodynamic mass a BH in GP theories decays into thermal radiation in the AdS spacetime of GR, while for a larger thermodynamic mass a BH peculiar to GP theories has a transition to a GR BH.
On the other hand, for $Y_0<0$, for a smaller thermodynamic mass a GR BH decays into the thermal radiation in the AdS spacetime peculiar to GP theories, while for a larger thermodynamic mass a GR BH theories has a transition to a BH peculiar to GP theories.
The ratio $S_{\rm GP}/S_{\rm GR}$ decreases as $Y_0$ increases, while increases as $Y_0$ decreases.

\subsection{RN-(A)dS solution with cubic-order GP interaction}
\label{sec5b}

We consider the class of  GP theories with generalized cubic-order GP interaction
\be
\label{generalized_cubic_model2}
G_2=\mF-\frac{\Lambda}{8\pi G},
\qquad 
G_3=g_3(Y),
\qquad 
G_4=\frac{1}{16\pi G},
\qquad 
c_2=0,
\ee
where $g_3(Y)$ represents the regular function of $Y$.
The theory \eqref{generalized_cubic_model2} admits the RN-(A)dS BH solution \cite{Heisenberg:2017xda,Heisenberg:2017hwb}
\be
\label{generalized_cubic_sol2}
h(r)
&=&
f(r)
=
-\frac{\Lambda r^2}{3}
+1
-
\frac{r_g}{r}
\left(
\frac{4\pi GQ^2}{r_g^2}
+1
-\frac{r_g^2\Lambda}{3}
\right)
+
\frac{4\pi GQ^2}{r^2},
\nonumber\\
A_0(r)
&=&
q+\frac{Q}{r},
\qquad 
A_1(r)
=
\frac{1}{f(r)}
\sqrt{A_0(r)^2-2Y_0 f(r)},
\ee
where $q$ and $Q$ are integration constants.
The existence of the solution requires that the norm of the vector field is given by the constant $Y=Y_0$ which satisfies
\be
\frac{\partial g_3}{\partial Y}
(Y_0)
=0.
\ee
This is the unique static and spherically symmetric BH solution with a nontrivial vector field in the given theory~\eqref{generalized_cubic_model2}.
The variation of the Hamiltonian on the horizon $r=r_g$ yields
\be
\label{var_horizon_3_21}
\delta{\cal H}_H
=
\frac{\delta r_g}{2G}
\left(
1
-
\Lambda r_g^2
-
\frac{4\pi  GQ^2}{r_g^2}
\right)
+
4\pi
A_0(r_g)
\delta Q.
\ee
The term proportional to $\delta q$ is canceled by the same contribution at the infinity $r\to\infty$ in the conservation of the total Hamiltonian $\delta {\cal H}=0$.
Thus, the differential of the BH entropy is given by 
$T_{\textsf{H}({\rm GP})} \delta S_{\rm GP}=\frac{\delta r_g}{2G}\left(1-\Lambda r_g^2-\frac{4\pi  GQ^2}{r_g^2}
\right)$,
where the Hawking temperature \eqref{hawking_temperature} is given by $T_{\textsf{H}({\rm GP})}=T_0\left(1-\Lambda r_g^2-\frac{4\pi  GQ^2}{r_g^2}\right)$.
Thus, by integrating this, we obtain the Bekenstein-Hawking entropy as the BH entropy $S_{\rm GP}=S_0$.

The thermodynamic mass of the BH is given by 
\be
M_{\rm GP}
=
M_0
\left(
1
+
\frac{4\pi  GQ^2}{r_g^2}
-
\frac{\Lambda r_g^2}{3}
\right),
\ee
which coincide with the ADM mass $M_{\rm ADM}=M_{\rm GP}$.
We have 
$
\frac{\partial M_{\rm GP}}{\partial r_g}
\delta r_g
=
\delta M_{\rm GP}
-
\frac{\partial M_{\rm GP}}
        {\partial Q}
\delta Q
=
\delta M_{\rm GP}
-
\Phi_H
\delta Q$,
where $\Phi_H:=-4\pi \left(A_0(r\to \infty)-A_0(r_g)\right)$.
The conservation of the Hamiltonian ${\cal H}=0$ for the solution \eqref{generalized_cubic_sol2}
yields the first law of BH thermodynamics in the presence of the electric charge $Q$
\be
\label{charged_first_law_cubic}
T_{\textsf{H}({\rm GP})} \delta S_{\rm GP}
=
\delta M_{\rm GP}
-
\Phi_{H}
\delta Q,
\ee
which coincides with the first law in EM theory \eqref{charged_first_law_em}.

In the same theory \eqref{generalized_cubic_model2}, there also exists the Schwarzschild-(A)dS solution in GR 
\be
\label{gr_rn}
h(r)
=
f(r)
=
-\frac{\Lambda r^2}{3}
+1
-
\frac{r_g}{r}
\left(
1
-\frac{r_g^2\Lambda}{3}
\right),
\quad 
A_0(r)=A_1(r)=0.
\ee

\subsection{RN-(A)dS solution with quadratic-order GP interaction}
\label{sec5c}

We consider the class of GP theories with  generalized quadratic-order GP interaction
\be
\label{generalized_quadratic_model2}
G_2=\left(1+2g_2(Y) \right)\mF-\frac{\Lambda}{8\pi G},
\qquad 
G_3=0,
\qquad 
G_4=\frac{1}{16\pi G},
\qquad 
c_2=0,
\ee
where $g_2(Y)$ represents the regular function of $Y$.
The theory \eqref{generalized_quadratic_model2} admits the RN-(A)dS BH solution \cite{Heisenberg:2017xda,Heisenberg:2017hwb}
\be
\label{generalized_quadratic_sol2}
h(r)
&=&
f(r)
=
-\frac{\Lambda r^2}{3}
+1
-
\frac{r_g}{r}
\left(
\frac{4\pi GQ^2}{r_g^2}
\left(
1+2g_2(Y_0)
\right)
+1
-\frac{r_g^2\Lambda}{3}
\right)
+
\frac{4\pi GQ^2}{r^2}
\left(
1+2g_2(Y_0)
\right),
\nonumber\\
A_0(r)
&=&
q+\frac{Q}{r},
\qquad 
A_1(r)
=
\frac{1}{f(r)}
\sqrt{A_0(r)^2-2Y_0 f(r)},
\ee
where $q$ and $Q$ are integration constants.
The existence of the solution requires that the norm of the vector field is given by the constant $Y=Y_0$ which satisfies
\be
\frac{\partial g_{2}}{\partial Y}(Y_0)=0.
\ee
This is the unique static and spherically symmetric BH solution with a nontrivial vector field in the given theory \eqref{generalized_quadratic_model2}.
We note that the same solution as Eq.~\eqref{generalized_quadratic_sol2} can be obtained from the equivalent description of the same theory as Eq.~\eqref{generalized_quadratic_model2}
\be
\label{generalized_quadratic_model2_alternative}
G_2=\mF-\frac{\Lambda}{8\pi G},
\qquad 
G_3=0,
\qquad 
G_4=\frac{1}{16\pi G}+\beta g_4(Y),
\qquad 
c_2=-\frac{g_2(Y)}{G_{4Y}},
\ee
by taking the limit of $\beta\to 0$ after the derivation of the EL equations.

The variation of the Hamiltonian on the horizon $r=r_g$ yields
\be
\label{var_horizon_3_22}
\delta{\cal H}_H
=
\frac{\delta r_g}{2G}
\left(
1
-
\Lambda r_g^2
-
\frac{4\pi  GQ^2}{r_g^2}
\left(
1+2g_2(Y_0)
\right)
\right)
+
4\pi
A_0(r_g)
\left(
1+2g_2(Y_0)
\right)
\delta Q.
\ee
The term proportional to $\delta Q$ in Eq.~\eqref{var_horizon_3_22} cancels the same contribution at the infinity $r\to\infty$ in the conservation of the total Hamiltonian $\delta {\cal H}=0$.
Thus, the differential of the BH entropy is given by 
\be
T_{\textsf{H}({\rm GP})} \delta S_{\rm GP}
=\frac{\delta r_g}{2G}
\left(1-\Lambda r_g^2-\frac{4\pi  GQ^2}{r_g^2}\left(1+2g_2(Y_0)\right)\right),
\ee
where the Hawking temperature \eqref{hawking_temperature} is given by
$T_{\textsf{H}({\rm GP})}=T_0\left[1-\Lambda r_g^2-\frac{4\pi  GQ^2}{r_g^2}\left(1+2g_{2}(Y_0)\right)\right]$.
Thus, by integrating this, we obtain the Bekenstein-Hawking entropy $S_{\rm GP}=S_0$.

The thermodynamic mass of the BH is given by 
\be
\label{bcl_mass_f}
M_{\rm GP}
=
M_0
\left(
1
-\frac{r_g^2\Lambda}{3}
+
\frac{4\pi GQ^2}{r_g^2}
\left(
1+2g_2(Y_0)
\right)
\right),
\ee
which coincide with the ADM mass $M_{\rm GP}=M_{\rm ADM}$.
We have 
$
\frac{\partial M_{\rm GP}}{\partial r_g}
\delta r_g
=
\delta M_{\rm GP}
-
\frac{\partial M_{\rm GP}}
        {\partial Q}
\delta Q
=
\delta M_{\rm GP}
-
\left(
1+2g_2(Y_0)
\right)
\Phi_H
\delta Q$,
where $\Phi_H:=-4\pi \left(A_0(r\to \infty)-A_0(r_g)\right)=\frac{4\pi Q}{r_g}$.
By defining the effective charge ${\tilde Q}:=\sqrt{1+2g_2(Y_0)}Q$ and the corresponding effective electric potential on the horizon ${\tilde \Phi}_H:=\frac{4\pi {\tilde Q}}{r_g}$, 
the first law of BH thermodynamics for the solution \eqref{generalized_quadratic_sol2} is given by 
\be
T_{\textsf{H}({\rm GP})} \delta S_{\rm GP}
=
\delta M_{\rm GP}
-
\tilde{\Phi}_H
\delta \tilde{Q}.
\ee
In the same theory \eqref{generalized_quadratic_model2}, there also exists the Schwarzschild-(A)dS solution with the trivial vector field as in GR, Eq.~\eqref{gr_rn}.

\section{Conclusion}
\label{sec6}

We have investigated thermodynamics of static and spherically symmetric BHs in GP theories. 
In order to obtain the variation of the total Hamiltonian, following the prescription by Iyer and Wald, we have employed the Noether charge associated with the diffeomorphism invariance.
The variation of the total Hamiltonian coincides with the variation of the Noether charge evaluated on the boundaries of the Cauchy surface.
In a BH system, the variations of the Hamiltonian on the event horizon and at the spatial infinity, respectively, give rise to the differentials of the entropy and the thermodynamic mass of the BH, respectively.
The conservation of the total Hamiltonian leads to the first law of BH thermodynamics.

Solutions in GP theories can be divided into the two classes.
The first class (Class I) corresponds to the solutions obtained by the direct promotion of the solutions in shift-symmetric Horndeski theories with the replacement of $\nabla_\mu \varphi\to A_\mu$.
The second one (Class II) consists of the solutions which are obtained only in GP theories and possess nonzero electromagnetic fields.
For the solutions in Class I, we have explicitly confirmed that the thermodynamic properties of the static and spherically symmetric BH solutions, i.e., the entropy and thermodynamic mass of BHs, remain the same as those in the counterpart solutions in shift-symmetric Horndeski theories, although BH thermodynamics has been constructed independently.
We have also discussed thermodynamic properties of static and spherically symmetric BHs in Class II.
In Tables \ref{table2} and \ref{table3}, we classify BH solutions in GP theories into Class I and II, and summarize their thermodynamical variables.
In Table \ref{table2}, we show the results of the Class I solutions.
Those BH solutions do not contain an electric field. 
Similarly, in Table \ref{table3}, we summarize thermodynamic properties of  the Class II solutions. 
Those BH solutions possess a nonzero electric field.

In Sec. \ref{sec4}, we have shown that thermodynamic variables of stealth Schwarzschild and Schwarzschild-(A)dS solutions in GP theories remain the same as those of the counterpart solutions in shift-symmetric Horndeski theories discussed in Ref.~\cite{Minamitsuji:2023nvh}.
For the same thermodynamic masses, the entropies of stealth Schwarzschild and Schwarzschild-(A)dS solutions have been investigated to study their thermodynamical stability in Secs. \ref{sec4a1} and \ref{sec4b1}, respectively.

As discussed in Sec.~\ref{sec4}, we have found that the entropy and the thermodynamic mass of the 
stealth Schwarzschild BH and the 
Schwarzschild-(A)dS BH with the electric field, which are obtained only for the specific value of the coupling constant of the vector field to the Einstein tensor, $(1/4)G^{\mu\nu}A_\mu A_\nu$, exactly coincide with those of the electrically neutral stealth Schwarzschild BH and the electrically neutral stealth Schwarzschild-dS BH in GP theories in the limit of the corresponding value of the coupling constant.
In other words, in these solutions the electric field does not affect thermodynamic properties of the BHs as well as the background BH solutions.
Two
entropies of 
stealth Schwarzschild solution and Schwarzschild-(A)dS solution with the electric field have been compared for stability analysis in Secs. \ref{sec4a2} and \ref{sec4b2}, respectively.
The same theory with this coupling constant also admitted the BH solution with $f(r)=h(r)=1-\sqrt{\frac{r_g}{r}}$, whose thermodynamic properties have been discussed in Sec. \ref{sec4a3}. Since this BH solution is not asymptotically flat and the thermodynamical transition to 
the stealth Schwarzschild solutions with the electric field is not possible, we did not discuss the stability by thermodynamical variables.

We have also investigated thermodynamic properties of other static and spherically symmetric BH solutions in GP theories in the presence of the generalized cubic- and quadratic-order GP interactions.
There exist the Schwarzschild-(A)dS and RN-(A)dS solutions with the constant norm of the vector field, respectively.
As discussed in Sec. \ref{sec5a},
in the case of the Schwarzschild-(A)dS solutions, we have shown that for the spacelike vector field the entropy of the GP BH is larger than that of the Schwarzschild-(A)dS BH in GR.
As discussed in Secs. \ref{sec5b} and \ref{sec5c}, in the case of the RN-(A)dS BH solutions, we have shown that thermodynamic properties of BHs remain the same as those in EM theory.

\begin{table}[htbp]
\begin{center}
\scalebox{0.9}[0.9]{
  \begin{tabular}{|c|c|c|c|c|c|}
\hline 
Class I&
action &
solution
&
mass
&
entropy
&
temperature
\\
\hline
&&&&&
\\[-.5em]
Ia& $G_2=\mF$, $G_4=\frac{1}{16\pi G}+\beta Y$,
&
stealth Schwarzschild
&
$\left(1-8\pi Gq^2\beta\right)M_0$
&
$\left(1-8\pi Gq^2\beta\right)S_0$
&
$T_{0}$
\\
&&&&&
\\[-.5em]
&$G_3=c_2=0$
&
GR Schwarzschild
&
$M_0$
&
$S_0$
&
$T_0$
\\
&&&&&
\\[-.5em]
\hline
Ib& 
$G_2=\mF+2m^2Y-\frac{\Lambda}{8\pi G}$,
&
GP Schwarzschild-(A)dS
&
$\left(1-\frac{r_g^2{\bar \Lambda}}{3}\right)M_0$
&
$\frac{\Lambda+\bar\Lambda}     {2\bar\Lambda}S_0$
&
$\left(1-\bar \Lambda r_g^2\right)T_0$
\\
&
$G_4=\frac{1}{16\pi G}+\beta Y$,
$G_3=c_2=0$
&
GR Schwarzschild-(A)dS
&
$\left(1-\frac{\Lambda r_g^2}{3}\right)M_0$
&
$S_0$
&
$\left(1-\Lambda r_g^2\right)T_{0}$
\\
\hline
 \end{tabular}
 }
\caption{Thermodynamic properties of BHs in GP theories which have counterpart solutions in the shift-symmetric Horndeski theories. $T_0$, $S_0$, and $M_0$ are defined in Eqs.~\eqref{T0}, \eqref{S0}, and \eqref{M0}, respectively. $q$ and ${\bar \Lambda}$ are defined in Eqs.~\eqref{solution_stealth} and \eqref{barLambda}, respectively.
The entropies of two BHs for the theories Ia and Ib are compared to study their thermodynamical stability in Sec.~\ref{sec4a1} and Sec.~\ref{sec4b1}, respectively.}
\label{table2}
\end{center}
\end{table}

\begin{table}[htbp]
\begin{center}
\scalebox{0.8}[0.9]{
  \begin{tabular}{|c|c|c|c|c|c|}
\hline 
Class II& action
&
solution
&
mass
&
entropy
&
temperature
\\
\hline
&&&&&
\\[-1em]
IIa&
$G_2=\mF$, $G_4=\frac{1}{16\pi G}+\frac{Y}{4}$,
&
stealth Schwarzschild
&
$\left(1-2\pi Gq^2\right)M_0$
&
$\left(1-2\pi Gq^2\right)S_0$
&
$T_{0}$
\\
&
$G_3=c_2=0$
&
GR Schwarzschild
&
$M_0$
&
$S_0$
&
$T_0$
\\
&
&
BH solution with $A_1=0$
&
$\frac{M_0}{2}$
&
$S_0$
&
$\frac{T_{0}}{2}$
\\
&&&&&
\\[-.7em]
\hline
IIb&
$G_2=\mF+2m^2Y-\frac{\Lambda}{8\pi G}$,
&
GP Schwarzschild-(A)dS
&
$\left(1-\frac{{\bar \Lambda} r_g^2}{3}\right)\frac{\Lambda+{\bar\Lambda}} {2\bar\Lambda}M_0$
&
$\frac{\Lambda+\bar\Lambda}     {2\bar\Lambda}S_0$
&
$\left(1-\bar \Lambda r_g^2\right)T_0$
\\
&
$G_4=\frac{1}{16\pi G}+\frac{Y}{4}$,
$G_3=c_2=0$
&
GR Schwarzschild-(A)dS
&
$\left(1-\frac{\Lambda r_g^2}{3}\right)M_0$
&
$S_0$
&
$\left(1-\Lambda r_g^2\right)T_{0}$
\\
&&&&&
\\[-.7em]
\hline
&&&&&
\\[-.5em]
IIc&
$G_2=g_2(Y)$, $G_3=g_3(Y)$
&
GP Schwarzschild-(A)dS
&
$\left(1+\frac{8\pi G}{3}r_g^2g_2(Y_0)\right)M_0$
&
$S_0$
&
$\left(1+8\pi G g_2(Y_0) r_g^2\right)T_0$
\\
&
$G_4=\frac{1}{16\pi G}$,
$c_2=0$
&
GR Schwarzschild-(A)dS
&
$\left(1-\frac{\Lambda r_g^2}{3}\right)M_0$
&
$S_0$
&
$\left(1-\Lambda r_g^2\right)T_{0}$
\\
\hline
IId&
$G_2=\mF-\frac{\Lambda}{8\pi G}$, $G_3=g_3(Y)$
&
GP RN-(A)dS
&
$\left(1+\frac{4\pi  GQ^2}{r_g^2}-\frac{\Lambda r_g^2}{3}\right)M_0$
&
$S_0$
&
$\left[1-\Lambda r_g^2-\frac{4\pi  GQ^2}{r_g^2}\right]T_0$
\\
&
$G_4=\frac{1}{16\pi G}$,
$c_2=0$
&
GR Schwarzschild-(A)dS
&
$\left(1-\frac{\Lambda r_g^2}{3}\right)M_0$
&
$S_0$
&
$\left(1-\Lambda r_g^2\right)T_0$
\\
\hline
IIe&
$G_2=\left(1+2g_2(Y) \right)\mF-\frac{\Lambda}{8\pi G}$
&
GP RN-(A)dS
&
$\left[
1
-\frac{r_g^2\Lambda}{3}
+
\frac{4\pi GQ^2}{r_g^2}
\left(
1+2g_2(Y_0)
\right)
\right]M_0$
&
$S_0$
&
$\left[1-\Lambda r_g^2-\frac{4\pi  GQ^2}{r_g^2}\left(1+2g_{2}(Y_0)\right)\right]T_0$
\\
&
$G_4=\frac{1}{16\pi G}$,
$G_3=c_2=0$
&
GR Schwarzschild-(A)dS
&
$\left(
1
-\frac{r_g^2\Lambda}{3}
\right)M_0$
&
$S_0$
&
$\left(1-\Lambda r_g^2\right)T_0$
\\
\hline
 \end{tabular}
 }
 \caption{Thermodynamic properties of BHs in GP theories which do not have counterpart solutions in the shift-symmetric Horndeski theories. $T_0$, $S_0$, and $M_0$ are defined in Eqs.~\eqref{T0}, \eqref{S0}, and \eqref{M0}, respectively.
 $q$ and ${\bar \Lambda}$ are defined in Eqs.~\eqref{solution_stealth2} and \eqref{barLambda2}, respectively.
 The entropies of two BHs for the theories IIa, IIb and IIc are compared to study their thermodynamical stability in  Secs.~\ref{sec4a2} and \ref{sec4a3}, 
Sec.~\ref{sec4b2}, and 
Sec.~\ref{sec5a}, respectively.}
 \label{table3}
\end{center}
\end{table}

As an extension of this work, it would be important to analyze thermodynamic properties of BH solutions in GP theories, which can be constructed only numerically.
It would also be interesting to extend our analysis of BH thermodynamics to solutions in vector-tensor theories beyond GP, especially stealth solutions and Schwarzschild-(A)dS solutions.
As in the Horndeski theories \cite{Hajian:2020dcq}, there could be potential ambiguities about the definition of the temperature of the temperature in the generalized Proca theories.
Since in the generalized Proca theories gravitons propagate with a speed which is different from the speed of light,
the black hole event horizon does not play the role as the causal boundary for gravitons.
In order to obtain an effective horizon for gravitons, a disformal transformation should be performed to the black hole solutions in the original generalized Proca theories.
The black hole temperature and thermodynamic quantitties computed for the effective horizions for gravitions could be different from those computed for the event horizon.
However, theories rewritten through such a disformal transformation does not belong to a class of the generalized Proca theories, and the formulation of black hole thermodynamics in such an extended theory is beyond the scope of this paper.
We hope to come back to these issues in our future work.

~~
\\

\section*{ACKNOWLEDGMENTS}
M.M.~was supported by the Portuguese national fund through the Funda\c{c}\~{a}o para a Ci\^encia e a Tecnologia in the scope of the framework of the Decree-Law 57/2016 of August 29, changed by Law 57/2017 of July 19, and the Centro de Astrof\'{\i}sica e Gravita\c c\~ao through the Project~No.~UIDB/00099/2020.
This work was supported in part by JSPS KAKENHI Grant Numbers  JP17H06359, JP19K03857.
 We would like to acknowledge the Yukawa Institute for Theoretical Physics at Kyoto University, where the present work has been completed during the long-term workshop ``Gravity and Cosmology 2024".

\appendix

\bibliography{refs}
\end{document}